\documentclass[review]{elsarticle}

\usepackage{lineno}
\modulolinenumbers[5]

\usepackage{mathptmx}      
\usepackage{latexsym}
\usepackage{graphicx}
\usepackage{amssymb}
\usepackage{amsmath}
\usepackage{times}
\usepackage{lipsum}
\usepackage{caption}
\usepackage{subcaption}
\usepackage{rotating}
\usepackage{xcolor}

\journal{Elsevier}

\begin{document}

\begin{frontmatter}


\title{Spatiotemporal pattern formation in nonlinear coupled reaction-diffusion systems with a mixed-type modal discontinuous Galerkin approach}

\author[address1]{Satyvir Singh \corref{cor1}}
\ead{satyvir.singh@ntu.edu.sg}
\cortext[cor1]{Corresponding author}

\author[address1]{Marco Battiato}

\author[address2]{Vinesh Kumar}

\address[address1]{School of Physical and Mathematical Sciences,
		Nanyang Technological University, 21 Nanyang Link, Singapore 637371}
	
\address[address2]{Department of Computer Science, Bharati College, University of Delhi, New Delhi 110054, India}


\begin{abstract}
The nonlinear coupled reaction-diffusion (NCRD) systems are important in the formation of spatiotemporal patterns in many scientific and engineering fields, including physical and chemical processes, biology, electrochemical processes, fractals, viscoelastic materials, porous media, and many others. In this study, a mixed-type modal discontinuous Galerkin approach is developed for one- and two- dimensional NCRD systems, including linear, Gray-Scott, Brusselator, isothermal chemical, and Schnakenberg models to yield the spatiotemporal patterns. These models essentially represent a variety of complicated natural spatiotemporal patterns such as spots, spot replication, stripes, hexagons, and so on. In this approach, a mixed-type formulation is presented to address the second-order derivatives emerging in the diffusion terms. For spatial discretization, hierarchical modal basis functions premised on the orthogonal scaled Legendre polynomials are used.  Moreover, a novel reaction term treatment is proposed for the NCRD systems, demonstrating an intrinsic feature of the new DG scheme and preventing erroneous solutions due to extremely nonlinear reaction terms. The proposed approach reduces the NCRD systems into a framework of ordinary differential equations in time, which are addressed by an explicit third-order TVD Runge-Kutta algorithm. The spatiotemporal patterns generated with the present approach are very comparable to those found in the literature. This approach can readily be expanded to handle large multi-dimensional problems that come up as model equations in developed biological and chemical applications. 
\end{abstract}

\begin{keyword}
Discontinuous Galerkin method \sep mixed-type formulation \sep NCRD systems \sep reaction term treatment \sep TVD Runge-Kutta algorithm \sep spatiotemporal patterns
\end{keyword}

\end{frontmatter}

\section{Introduction}
\label{Sec:1}

Over last several decades, the studies of spatiotemporal pattern formation in nature have been a fascinating subject for different areas of scientific research. These research areas include biology \cite{Murray1989Biology}, ecology \cite{segel1972dissipative}, semiconductor physics \cite{balkarei1988regenerative}, material science \cite{Krinsky1984Equilibrium}, hydrodynamics \cite{white1988planforms}, astrophysics \cite{nozakura1984formation}, and many more. 
In 1952, mathematician Alan Turing \cite{turing1952chemical} developed a nonlinear coupled reaction-diffusion (NCRD) systems based theory that featured an activator-inhibitor pair to describe the early development of spatiotemporal patterns in an embryo. In this hypothesis, he looked at how the activator's and inhibitor's diffusivities differ, leading to spontaneous symmetry breaking and spatially different stable patterns. After then, different type of Turing-type spatiotemporal pattern formation has been widely discovered in biology, chemistry, and population dynamics. Typical examples of biological applications to the spatiotemporal patterns are: hydra skin patterns \cite{gierer1972theory}, animal coat patterns \cite{murray1981pattern}, butterfly wing patterns \cite{nijhout1990comprehensive}, zebra coat patterns \cite{bard1981model}, skeletal patterning in limb evolution \cite{maini1991cellular} and sea shell coloring patterns \cite{Meinhardt1995Shells}. Several mathematical models based on spatiotemporal patterns utilized in biology, ecology, and biochemistry involve organisms reacting in the existence of diffusion, resulting in the NCRD systems. Few examples of well-known NCRD models are the Gierer and Meinhardt \cite{gierer1972theory}, the Brusselator \cite{prigogine1968symmetry}, the Schnakenberg \cite{schnakenberg1979simple}, the isothermal \cite{merkin1990development}, and the Gray-Scott \cite{gray1984autocatalytic}.

The scope of this research is limited to a two-dimensional NCRD systems with two chemical species which are stated in a generic form as
\begin{equation}
	\label{Eq:1}
	\frac{\partial \mathbf{U}}{\partial t} = \mathbf{D} \nabla^{2} \mathbf{U} + \mathbf{S}(\mathbf{U}),  \quad  \forall \, (\mathbf{x},t) \in \Omega \times [0,T],
\end{equation}
with initial conditions:
\begin{equation}
	\label{Eq:2}
	\mathbf{U}(\mathbf{x},0) = \mathbf{U}_{0}(\mathbf{x}), \quad \forall \, (\mathbf{x}) \in \Omega,
\end{equation}
and Dirichlet or Neumann boundary conditions. In above-mentioned expression,
\begin{equation}
	\label{Eq:3}
	\mathbf{U} = \begin{pmatrix}
		u \\ v
	\end{pmatrix}, \quad
	\mathbf{D}	= 
	\begin{pmatrix}
		\mu_{1} & 0\\ 
		0 & \mu_{2}
	\end{pmatrix}, \quad
	\mathbf{S}(\mathbf{U}) =
	\begin{pmatrix} 
		f(u,v)\\ 
		g(u,v)
	\end{pmatrix}, \quad
	\text{and} \quad \mathbf{x} = (x,y) \in \Omega,
\end{equation}
Here, $\mathbf{U}$ is the unknown vector of two chemical-species component $u,v$. $\mathbf{D}$ is the diffusion constant matrix with $\mu_{1}, \mu_{2}$ parameters. $\nabla^{2}$ is the Laplacian operator. While, $\mathbf{S}(\mathbf{U})$ represents the reaction matrix of two nonlinear reaction kinetics $f(u,v)$ and $g(u,v)$. The most general form of these reaction kinetics might be represented as
\begin{equation}
	\label{Eq:4}
	\begin{aligned}
		f(u,v) & = a_{1}u^{2}v + a_{2}uv^{2} + a_{3}uv + a_{4}u + a_{5}v + a_{6}, \\
		g(u,v) & = b_{1}u^{2}v + b_{2}uv^{2} + b_{3}uv + b_{4}u + b_{5}v + b_{6},
	\end{aligned}
\end{equation}
where $a_{i}$, $b_{i}$, $i=1,6$ represent the genuine parameters. 


To understand and predict the complex spatiotemporal dynamical structures, including the traveling wave and self-organized patterns found in nature such as spots, spot replication, stripes, hexagons and other dissipative solitons \cite{gray1984autocatalytic,pearson1993complex,rodrigo2001exact}, the mathematical modeling and numerical simulation of the NCRD systems are essential. The numerical solutions of these systems have been a major and challenging research area due to the stiff diffusion term paired with the highly nonlinear reaction term. When performing parametric and numerical investigations in the computational biology, a practical and reliable numerical method to solve the NCRD systems is required. 

Various studies have been carried out over the last few decades in attempt to develop numerical approaches for accurately approximating solutions of the NCRD systems.
Ruuth \cite{ruuth1995implicit} used implicit-explicit schemes in conjunction with spectral methods to describe the spatiotemporal patterns in biological and chemical processes. 
Zegeling and Kok \cite{zegeling2004adaptive} employed an adaptive moving mesh approach for the spatiotemporal patterns of one and two-dimensional NCRD systems with chemistry applications. Lopez and Ramos \cite{garcia1996linearized} used linearized methods in conjunction with operator-splitting techniques to obtain the solutions of isothermal NCRD system. Madzvamuse \cite{madzvamuse2003moving,madzvamuse2006time} employed the moving grid based finite element approaches to derive the spatiotemporal patterns of the NCRD systems on fixed and continually increasing domains using time-stepping procedures. Mittal and Rohila \cite{mittal2016numerical} adopted the modified cubic B-spline based differential quadrature approach for obtaining the spatiotemporal patterns of various one-dimensional NCRD system. Subsequently, Jiwari et al. \cite{jiwari2017numerical} captured the different types of spatiotemporal patterns in several NCRD system through the modified trigonometric cubic B-spline functions, which are based on the differential quadrature approach. Later, Yadav and Jiwari \cite{yadav2019finite} illustrated a finite element scheme to investigate the computation simulations for the spatiotemporal patterns of the NCRD systems. Tamsir and co-authors \cite{tamsir2018cubic, dhiman2018collocation} proposed a hybrid numerical method of cubic trigonometric B-spline base functions and differential quadrature algorithm for solving Fisher type NCRD systems.
Onarcan et al. \cite{onarcan2018trigonometric} presented a trigonometric cubic B-splines based collocation scheme for obtaining the spatiotemporal dynamics of the NCRD systems. So far, some of the most well-known computational algorithms for developing the biological based spatiotemporal patterns are also carried out in the literature, for example, finite difference \cite{garvie2007finite}, positive finite volume \cite{gerisch2006robust}, 
finite volume spectral element \cite{shakeri2011finite}, and
meshless local Petrov–Galerkin \cite{ilati2017application} methods.

In this study, the authors adopted the discontinuous Galekrin (DG) technique to investigate the spatiotemporal pattern formation in the NCRD systems. Reed and Hill \cite{Reed1973DG} in 1973  established the DG technique for addressing neutron transport, and it has since become a popular tool for approximating a wide range of partial differential equations. For the numerical approximations as well as the test functions, this method uses a totally discontinuous piece-wise polynomial space. Over past years, Cockburn and co-authors devised a framework for solving PDEs based hyperbolic conservation laws in a series of works \cite{cockburn1989tvbII,cockburn1989tvb, cockburn1990runge, cockburn1991runge,cockburn1998local,cockburn1998runge,cockburn2001runge,cockburn2001runge}  that contributed significantly to the development of the DG technique. This method has been effectively implemented to several problems, including fluid flows, multi-phase flows, quantum physics, magneto-hydrodynamics, and many others \cite{franquet2012runge,madaule2014energy,filbet2018hybrid,raj2017super,Singh2020IACM,Singh2020Material,Singh2021CF,singh2021IJHMT,Singh2021JCP}, it has just recently made its way into developmental biology. This implementation is motivated by advancements in the approach as well as recent breakthroughs in spatiotemporal pattern generation, which have made the DG method a viable instrument for a broader range of biological and chemical requisitions.

Zhu et al. \cite{zhu2009application} merged the DG approach \cite{cheng2008discontinuous} along with the Strang type symmetrical operator splitting methodology to address the NCRD system. Subsequently, this technique has been employed to the moving meshes \cite{zhu2009finite}. Zhang et. al. \cite{zhang2014direct} provided a direct DG method, which is attributed to the direct weak formulation of the NCRD system and the construction of appropriate numerical flux scheme on the elemental boundaries. Later, Zhang et al. \cite{zhang2016new} illustrated a newly nonlinear Galerkin approach subjected to finite element discretization for the numerical approximations of the NCRD systems on wide intervals. Recently, Singh \cite{Singh2021IACM} presented a different scheme, so called mixed-type modal DG method to solve the generic FitzHugh–Nagumo RD equation. Later, this work was extended to the nonlinear RD Fisher-KPP equation \cite{Singh2022CRC}. 

\begin{figure}
	\centering
	\includegraphics[width=1.0\linewidth]{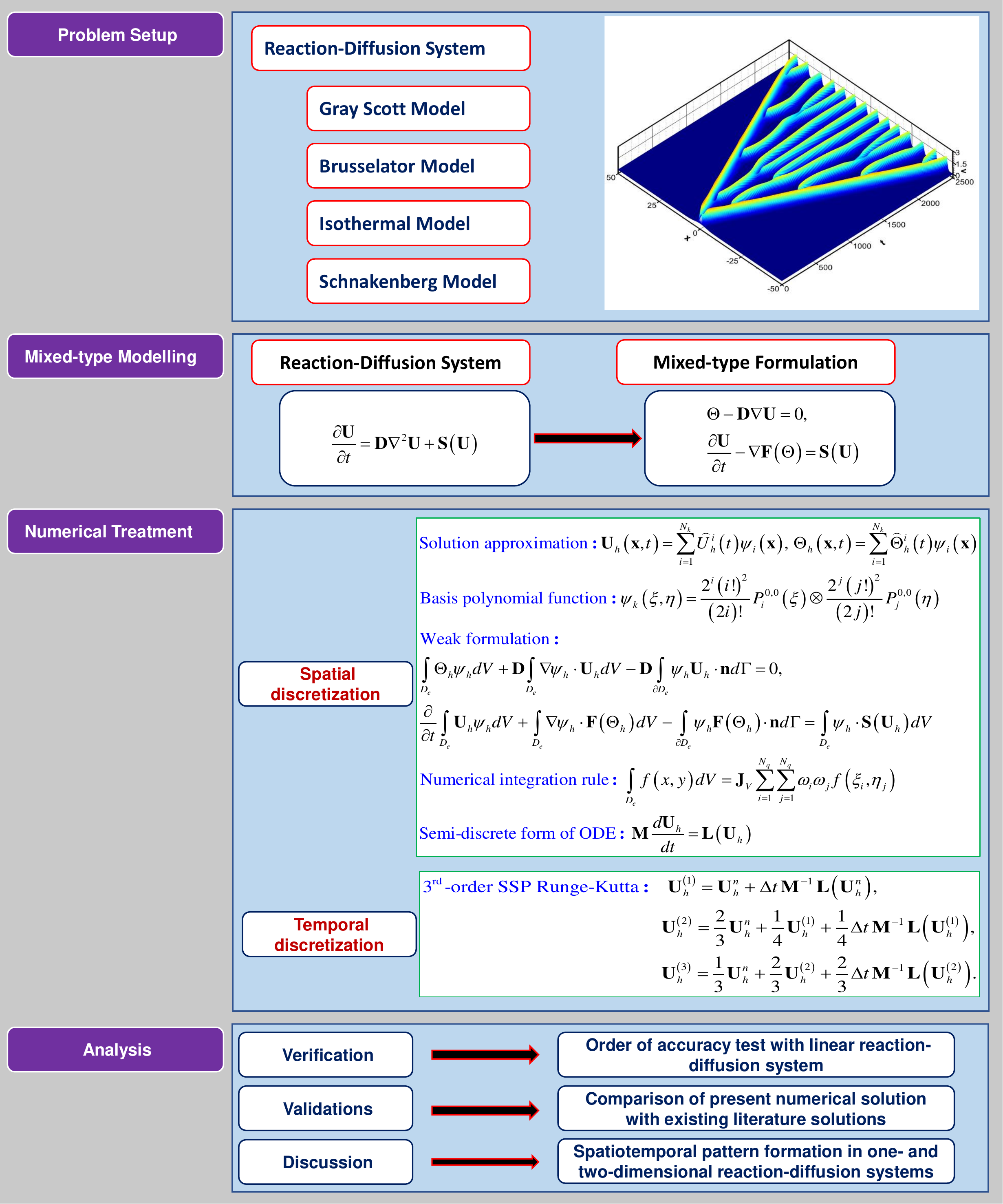}
	\caption{Graphical abstract of mixed-type modal discontinuous Galerkin approach.}
	\label{Fig:1}
\end{figure}

The aim of this study is to describe a mixed-type modal DG scheme for solving the NCRD systems (\ref{Eq:1}) on Cartesian grids that is an efficient and simple to implement. A graphical abstract for the proposed mixed-type modal DG approach is illustrated in Fig. \ref{Fig:1}. The following are the primary aspects of the proposed scheme:

$\bullet$ A mixed-type DG formulation is presented to address the second-order derivatives emerging in the diffusion terms.

$\bullet$ For the spatial discretization, hierarchical modal ansatz functions premised on the orthogonal scaled Legendre polynomials are used, which differs from earlier work on the NCRD systems.  

$\bullet$ A novel reaction term treatment is adopted for the NCRD systems, demonstrating an intrinsic feature of the new DG scheme and preventing erroneous solutions due to extremely nonlinear reaction terms.
  
$\bullet$  Finally, the proposed technique is innovative for NCRD systems, specifically Gray-Scott, Brusselator, Schnakenberg, isothermal, and linear models. Also, the captured spatiotemporal pattern resembles the models' existing patterns.

The outlines of the current paper are defined as: In Section \ref{Sec:2}, the mixed-type DG spatial discretization in detail is reported to solve the NCRD system (\ref{Eq:1}) in one-dimensional and two-dimensional space with Cartesian meshes. In Section \ref{Sec:3}, we demonstrate the accuracy and validation of the proposed numerical approach for linear RD model with exact solutions. The proposed scheme is then employed to one- and two-dimensional NCRD models utilized in developed biology to generate spatiotemporal patterns. We demonstrate the influence of changing parameters on the resulting patterns through various simulations of these models. Finally, in Section \ref{Sec:4}, a conclusion is offered, along with recommendations for further research in light of the current study. 

\section{Mixed-type modal discontinuous Galerkin method}
\label{Sec:2}

We present here a mixed-type modal discontinuous Galerkin (DG) formulation to solve the one and two-dimensional NCRD systems (\ref{Eq:1}),  which has been proven to be an appropriate technique for dealing with second-order derivatives in diffusion terms \cite{raj2017super,singh2021IJHMT,Singh2021JCP}. A new variable, $\Theta$, is introduced in this formulation, which can be thought of as the derivatives of unknown variables.
As a result, for the mixed-type DG construction, the NCRD system (\ref{Eq:1}) is regarded a connected framework of  $\mathbf{U}$ and $\Theta$.
\begin{equation}
	\label{Eq:5}
	\begin{aligned}
		\Theta - \mathbf{D} \, \nabla \mathbf{U} & = 0, \\
		\frac{\partial \mathbf{U}}{\partial t}  - \nabla \mathbf{F}(\Theta)  &  = \mathbf{S}(\mathbf{U}),
	\end{aligned}
\end{equation}
where $\mathbf{F}(\nabla \mathbf{U})= \nabla \mathbf{U}$ is the diffusive term.

\subsection{Mesh descriptions and modal basis functions}
\label{Sec:3.1}

\subsubsection{One-dimensional case}
\label{Sec:3.1.1}

\begin{figure}
	\centering
	\includegraphics[width=1.0\linewidth]{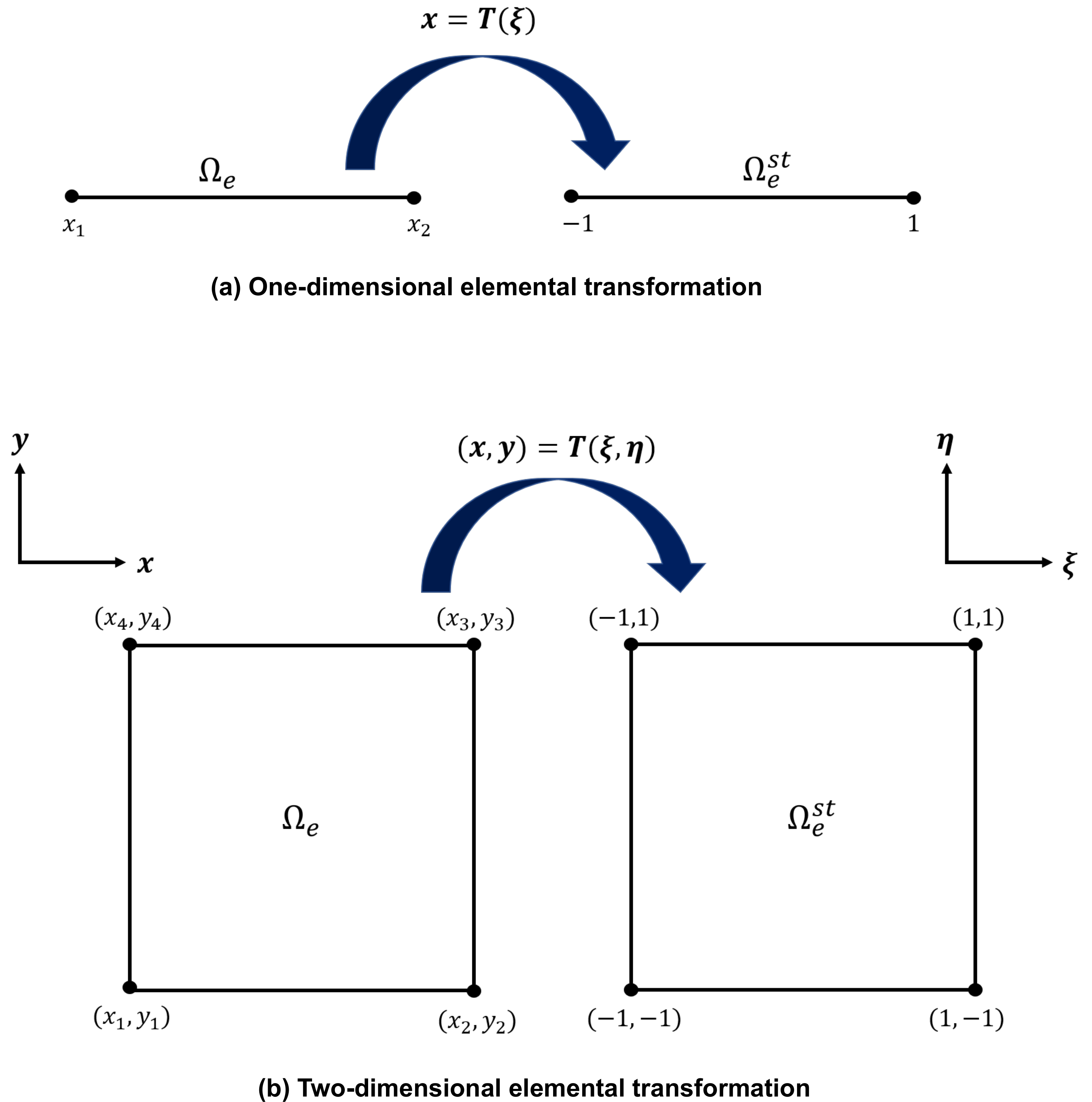}
	\caption{Transformation between the physical and the standard elements in (a) one-dimensional, and (b) two-dimensional cases.}
	\label{Fig:2}
\end{figure}

For one-dimensional spatial discretization of the NCRD systems (\ref{Eq:5}), the computational domain $D = [x_{L},x_{R}]$ is distributed into $N$ regular cells as follows:
\begin{equation}
	\label{Eq:6}
	x_{L} = x_{1/2} < x_{3/2} < \cdots < x_{N + 1/2} = x_{R}.
\end{equation}
The cell can be specified by $I_{n}=[x_{n - 1/2}, x_{n + 1/2}],  \quad  \forall
n\in 1,2, \cdots N$ with cell center $x_{n} = \frac{1}{2}(x_{n - 1/2} + x_{n +1/2})$ and cell size $\Delta x = x_{n + 1/2} - x_{n - 1/2}$ of the element $I_{n}$. 
The piece-wise polynomial space of the functions
$\upsilon_{h}:D \longmapsto \Re$ is addressed for the domain $D$ as
\begin{equation}
	\label{Eq:7}
	V_{h}^l =  \{ \upsilon_{h} \in L_2(D), \upsilon_{h} \,|\,_{D} \in \mathbb{P}^{l}(I_{n}), \quad  \forall n\in 1,2, \cdots N \}.
\end{equation}
Here, $L_2(D)$ represents the squared Lebesque integral over $D$, and $\mathbb{P}^{l}(I_{n})$ represents the space of polynomial functions of degree at most $l$ in $I_{n}$. Then, the true solutions of $\mathbf{U}$ and $\Theta$ can be approximated with $\mathbf{U}_h \in V_{h}^l$ and $\Theta_h \in V_{h}^l$, respectively.
\begin{equation}
	\label{Eq:8}
	\begin{aligned}
		\mathbf{U}_{h}(x,t) & = \sum_{i=1}^{N_{l}} \hat{{U}}^{i}_{h}(t)\psi_{i}(x),\\
		\Theta_{h}(x,t) & = \sum_{i=1}^{N_{l}} \hat{\Theta}^{i}_{h}(t)\psi_{i}(x), 
		\quad \forall (x)\in I_{n},
	\end{aligned}
\end{equation}
where $\hat{U}^{i}_{h}$ and $\hat{\Theta}^{i}_{h}$ are the modal coefficient for $\mathbf{U}$ and $\Theta$ to be enriched with time, and $\psi$ represents the basis function of degree $l$. In $\mathbb{P}^{l}(I_{n})$, $N_{l}=l+1$ is used to calculate the number of modal coefficients. This study deals with the hierarchical modal basis functions premised on the orthogonal scaled Legendre polynomials for the $\psi$ construction \cite{Signh2018Thesis}.
\begin{equation}
	\label{Eq:9}
	\psi_n(\xi)  = \frac{2^n (n!)^2}{(2n)!} P_{n}^{0,0}(\xi),
	\quad -1\leq \xi \leq 1, \quad  n \geq 0, \\
\end{equation}
where $P_{n}^{0,0}(\xi)$ represents the Legendre polynomial functions. For $\mathbb{P}^{3}$ case, which is related to the third-order approximations, the first four basis functions are calculated as
\begin{equation}
	\label{Eq:10}
	\psi_{0}(\xi)  = 1, \quad \psi_{1}(\xi) = \xi, \quad \psi_{2}(\xi) = \xi^{2} -\frac{1}{2}, \quad \psi_{3}(\xi) = \xi^{3} -\frac{3}{5} \xi. 
\end{equation}
Here, it should be noted that the standard local element $(D_{e}^{st})$ is mapped from the computational space $\xi\in [-1,1]$ to an arbitrary line element $(D_{e})$ under a linear mapping $T:D_{e}^{st} \longmapsto D_{e}$, as depicted in Fig. \ref{Fig:2}(a).
\begin{equation}
	\label{Eq:11}
	x  = \frac{1}{2}\left [ (1-\xi)x_{1} + (1+\xi)x_{2}     \right], \\
\end{equation}
where $x_{i}$, $i=1,2$ indicate the vertices of physical element $\Omega_{e}$. 

\subsubsection{Two-dimensional case}
\label{Sec:3.1.2}

For two-dimensional NCRD systems, the computational domain is partitioned with rectangular elements. Let us suppose that $\Im$ be a collection of segments of the computational domain $D = D_{x} \times D_{y}$, which is subdivided into $N_{x} \times N_{y}$ non-overlapping rectangular elements. A two-dimensional Cartesian element $\Im$ is given as
\begin{equation}
	\label{Eq:12}
	\Im:= \{T_{ij} = I_{i} \times J_{j}, \quad  \forall \, i \in  N_{x}, \, j \in  N_{y}  \},
\end{equation}
where 
\begin{equation}
	\label{Eq:13}
	\begin{aligned}
		I_{i}: & = \left [x_{i-1/2}, x_{i+1/2}  \right],  \quad \forall \, i \in  N_{x}, \\
		I_{j}: & = \left [y_{j-1/2}, y_{j+1/2}  \right],  \quad \forall \, j \in  N_{y},\\
	\end{aligned}
\end{equation}
where $[x_{i},y_{j}] = \frac{1}{2}[x_{i-1/2} + x_{i+1/2},y_{j-1/2} + y_{j+1/2}]$ is the center point of the rectangular element. For the $D_{x}$ domain, a set of piece-wise polynomial functions $\upsilon_{h}:D_{x}\longmapsto \Re$ can be offered as
\begin{equation}
	\label{Eq:14}
	Z_{h}^{l} =  \{ \upsilon_{h} \in L_2(D_{x}) :  \upsilon_{h} \,|\,_{D_{x}} \in \mathbb{P}^{l}(I_{i}), \quad  \forall \, i \in  N_{x} \},
\end{equation}
On the other hand, we can consider $\upsilon_{h}:D \longmapsto \Re$ for the domain $D =D_{x} \times D_{y}$ such as,
\begin{equation}
	\label{Eq:15}
	V_{h}^{l} =  \{\upsilon_{h} \in L_2(D) : \upsilon_{h} |\,_{D} \in \mathbb{Q}^{l}(T_{ij}), \quad \forall  \, i \in N_{x}, \,  j \in N_{y} \},
\end{equation}
where $\mathbb{Q}^{l}(T_{ij})=\mathbb{P}^{l}(I_{i}) \otimes \mathbb{P}^{l}(J_{j})$ is the set of polynomial functions with degree of at most $l$ on $T_{ij}$. The maximum number of modal coefficients for $\mathbb{Q}^{l}(T_{ij})$ is calculated by $N_{l}=(l+1)(l+2)/2$. 
Now, the true solutions of $\mathbf{U}$ and $\Theta$ can be estimated with $\mathbf{U}_h \in V_{h}^l$ and $\Theta_h \in V_{h}^l$, respectively, as
\begin{equation}
	\label{Eq:16}
	\begin{aligned}
		\mathbf{U}_{h}(x,y,t) & = \sum_{i=1}^{N_{k}} \hat{{U}}^{i}_{h}(t)\psi_{i}(x,y),\\
		\Theta_{h}(x,y,t) & = \sum_{i=1}^{N_{k}} \hat{\Theta}^{i}_{h}(t)\psi_{i}(x,y), 
		\quad (x,y)\in T_{ij}.
	\end{aligned}
\end{equation}
For the two-dimensional problems, the hierarchical modal basis functions premised on the scaled Legendre polynomials are utilized for $\varphi$, which are computed through the following mathematical expression \cite{Signh2018Thesis}
\begin{equation}
	\label{Eq:17}
	\psi_{k}(\xi, \eta) =  \frac{2^{i}(i!)^{2}}{(2i)!} P_{i}^{0,0}(\xi)    \otimes   \frac{2^{j}(j!)^{2}}{(2j)!} P_{j}^{0,0}(\eta),   \quad -1 \leq \xi, \eta  \leq 1
\end{equation}
The first six basis functions for $\mathbb{Q}^{2}$ case, which correspond to the third-order approximation, are calculated as
\begin{equation}
	\label{Eq:18}
	\psi_{1} = 1, \quad  \psi_{2} = \xi, \quad  \psi_{3} = \eta,  \quad  \psi_{4} = \xi^{2} - \frac{1}{2}, \quad \psi_{5} = \xi \eta, \quad  \psi_{6} = \eta^{2} - \frac{1}{2}.
\end{equation}
For these basis functions, a local Cartesian coordinate system $(\xi,\eta)\in [-1,1]$ is used for a reference rectangular element, which is mapped from the physical space $(x,y)$ under a linear mapping $T:D_{e}^{st} \longmapsto D_{e}$, as illustrated in Fig. \ref{Fig:2}(b).
\begin{equation}
	\label{Eq:19}
	\begin{aligned}
		x  &  = \frac{1}{2}\left [ (1-\xi)x_{1} + (1+\xi)x_{2}     \right], \\
		y  &  = \frac{1}{2}\left [ (1-\eta)y_{1} + (1+\eta)y_{2}    \right].
	\end{aligned}
\end{equation}
Here, $(x_{i},y_{i}), \ \forall 1 \leq i \leq 4$ denote the physical coordinates of the vertices of $D_{e}$. For such mapping, the Jacobian can be calculated as
\begin{equation}
	\label{Eq:20}    
	J_{x\rightarrow \xi} = \frac{\partial(x,y)}{(\xi,\eta)} = \frac{1}{4} 
	\begin{vmatrix}
		x_{2}-x_{1}      &  0\\
		0 &   y_{2}-y_{1}   \\
	\end{vmatrix}
	=\frac{1}{4} A_{e}.
\end{equation}
where rectangle $A_{e}$ represents the area of rectangle element $D_{e}$.

\subsection{Spatial discretization in the DG framework}
\label{Sec:3.2}

To spatial discretize in the DG method, we establish the following weak formulation for $\mathbf{U}_h \in V_{h}^{l}(\Im)$ and $\Theta_h \in V_{h}^{l}(\Im)$ of the NCRD system (\ref{Eq:4}), which is obtained by multiplying Eqs. (\ref{Eq:7}) and (\ref{Eq:15}) by a test (basis) function $\psi_{h}$, integrating over the domain $D_{e} \in \Im$, and executing the integration by parts:
\begin{equation}
	\label{Eq:21}
	\begin{aligned}
		\int \limits_{D_{e}} \Theta_{h} \psi_{h} dV + \mathbf{D}
		\int \limits_{D_{e}}  \nabla \psi_{h} \cdot \mathbf{U}_{h}  dV - \mathbf{D}
		\int \limits_{\partial D_{e}} \psi_{h} \mathbf{U}_{h} \cdot \mathbf{n} d\Gamma & = 0, \\
		\frac{\partial}{\partial t} \int \limits_{D_{e}} \mathbf{U}_{h} \psi_{h} dV 
		+ \int \limits_{D_{e}} \nabla \psi_{h} \cdot \mathbf{F}(\Theta_{h}) dV -
		\int \limits_{\partial D_{e}} \psi_{h} \mathbf{F}(\Theta_{h}) \cdot \mathbf{n} d\Gamma & = \int \limits_{D_{e}} \psi_{h} \mathbf{S}(\mathbf{U}_{h}) dV.
	\end{aligned}
\end{equation}
Here, $\mathbf{n}$ denotes the outward unit normal vector, while $V$ and $\Gamma$ denote the volume and boundary of the cell $D_{e}$, respectively. The numerical solutions $\mathbf{U}_{h}$ and $\Theta_{h}$ are discontinuous across element interfaces, therefore the interface fluxes are not uniquely defined. The flux functions $\mathbf{U}_{h} \cdot \mathbf{n}$, and $\mathbf{F} (\Theta_{h}) \cdot \mathbf{n}$  emerging in Eq. (\ref{Eq:21}) are substituted by a numerical flux function illustrated at the cell interfaces, which are denoted by $\mathbf{H}^{aux}$, and $\mathbf{H}^{alt}$, respectively. We used here an alternating scheme \cite{cockburn1998local} for the viscous and auxiliary flux functions, illustrated as 
\begin{equation}
	\label{Eq:22}
	\begin{aligned}
		\mathbf{U}_{h} \cdot \mathbf{n} & \equiv \mathbf{H}^{aux} (\mathbf{U}_{h}^{int},\mathbf{U}_{h}^{ext}) = \frac{1}{2} \left[\mathbf{U}_{h}^{int} + \mathbf{U}_{h}^{ext} \right] + C_{11} \left[\mathbf{U}_{h}^{int} - \mathbf{U}_{h}^{ext} \right], \\
		\mathbf{F}(\Theta_{h}) \cdot \mathbf{n} & \equiv \mathbf{H}^{alt} (\Theta_{h}^{int},\Theta_{h}^{ext}) =\frac{1}{2} \left[\mathbf{F}(\Theta_{h}^{int})+ \mathbf{F}(\Theta_{h}^{ext}) \right]
		- C_{11} \left[\mathbf{U}_{h}^{int} - \mathbf{U}_{h}^{ext} \right]
		+ C_{12} \left[\Theta_{h}^{int} - \Theta_{h}^{ext} \right],  
	\end{aligned}
\end{equation}
where the superscripts ``int" and ``ext" convey the interior and exterior states of the elemental interface. When $C_{11}=0$ and $C_{12}=1/2$ or $-1/2$, the fluxes are called the alternating fluxes \cite{cockburn1998local}. It can be noted that the mentioned flux scheme becomes central flux with with $C_{11}=C_{12}=0$. Thus, the aforementioned weak formulation (\ref{Eq:21}) can be expressed as follows:
\begin{equation}
	\label{Eq:23}
	\begin{aligned}
		\int \limits_{D_{e}} \Theta_{h} \psi_{h} dV +  \mathbf{D}
		\int \limits_{D_{e}}  \nabla \psi_{h} \cdot \mathbf{U}_{h}  dV -  \mathbf{D}
		\int \limits_{\partial D_{e}} \psi_{h} \mathbf{H}^{aux} d\Gamma & = 0, \\
		\frac{\partial}{\partial t} \int \limits_{D_{e}} \mathbf{U}_{h} \psi_{h} dV 
		+ \int \limits_{D_{e}} \nabla \psi_{h} \cdot \mathbf{F} dV -
		\int \limits_{\partial D_{e}} \psi_{h} \mathbf{H}^{alt} d\Gamma
		& = \int \limits_{D_{e}} \psi_{h} \mathbf{S}(\mathbf{U}_{h}) dV .
	\end{aligned}
\end{equation}

\subsection{Novel reaction terms treatment}
\label{Sec:3.3}

It is generally known that strictly hyperbolic systems with reaction terms have a wide range of relaxation times, which causes significant numerical challenges. Furthermore, when the reaction terms are not adequately resolved, spurious solutions might occur. As a result, it must be appropriately treated. Here, we show how the intrinsic feature of the novel DG scheme  eliminates the necessity for inefficient reaction term treatments in traditional approaches. Consider the reaction terms associated to the elemental formulation (\ref{Eq:23})  for a single variable $u_{h}$.
\begin{equation}
	\label{Eq:24}
	\frac{d}{d t} \int \limits_{D_{e}} u_{h} \psi_{h} dV 
	+ \int \limits_{D_{e}} \nabla \psi_{h} \cdot \mathbf{F} dV -
	\int \limits_{\partial D_{e}} \psi_{h} \mathbf{H}^{alt} d\Gamma
	= \int \limits_{D_{e}} \psi_{h} \mathbf{S}(u_{h}) dV .
\end{equation}
The above mentioned equation can also be represented in a matrix format by fascinating $\mathbf{U}$ as the global vector of modal coefficients:
\begin{equation}
	\label{Eq:25}
	\mathbf{M} \frac{d \mathbf{U}}{d t} + \mathbf{KU} - \mathbf{H}^{alt} (u_{h})  \Theta -  \mathbf{S}(u_{h}) \Theta^{'} =0.
\end{equation}
After some calculations, this equation may be read as
\begin{equation}
	\label{Eq:26}
	\begin{aligned}
	\frac{d \mathbf{U}}{d t} & = L(\mathbf{U}), \\
	\mathbf{U} & = (U^{(1)},U^{(2)},U^{(3)}, \dots U^{(N)})^{T}, \\
	L(\mathbf{U}) & = \mathbf{M}^{-1}  \mathbf{KU} + \mathbf{H}^{alt} (u_{h}) \mathbf{M}^{-1} \Theta -  \mathbf{S}(u_{h}) \mathbf{M}^{-1} \Theta^{'},
    \end{aligned}
\end{equation}
where $\mathbf{M}$ is the mass matrix, $\mathbf{K}$ is the stiffness matrix, $\Theta$ is the vector associated with the boundary term, and $\Theta^{'}$ is the vector associated with the source terms. The following are the definitions for these matrices:
\begin{equation}
	\label{Eq:27}
	\begin{aligned}
		\mathbf{M} & = \int \limits_{D_{e}} \psi_{i} \psi_{j} dV  \quad \forall \, \, 1 \leq i \leq j \leq N, \\
		& = \begin{bmatrix}
			\int \limits_{D_{e}} \psi_{1} \psi_{1} dV & \int \limits_{D_{e}} \psi_{1} \psi_{2} dV & \dots & \int \limits_{D_{e}} \psi_{1} \psi_{N} dV \\
			\int \limits_{D_{e}} \psi_{2} \psi_{1} dV & \int \limits_{D_{e}} \psi_{2} \psi_{2} dV & \dots & \int \limits_{D_{e}} \psi_{2} \psi_{N} dV \\
			\vdots & \vdots & \ddots & \vdots \\
			\int \limits_{D_{e}} \psi_{N} \psi_{1} dV & \int \limits_{D_{e}} \psi_{N} \psi_{2} dV & \dots & \int \limits_{D_{e}} \psi_{N} \psi_{N} dV \\
		\end{bmatrix}
	\end{aligned}
\end{equation}
Because of the orthogonal nature of the basic functions,
\begin{equation*}
	\mathbf{M} = \begin{cases}
		C_{ij}    \quad \, \, i = j \\
		0        \quad \quad i \neq j \\
	\end{cases},
\end{equation*}
\begin{equation}
	\label{Eq:28}
	\begin{aligned}
		\mathbf{K} & = \int \limits_{D_{e}} \nabla \psi_{i} \psi_{j} dV  \quad \forall \, \, 1 \leq i \leq j \leq N, \\
		& = \begin{bmatrix}
			\int \limits_{D_{e}} \nabla \psi_{1} \psi_{1} dV & \int \limits_{D_{e}} \nabla \psi_{1} \psi_{2} dV & \dots & \int \limits_{D_{e}} \nabla \psi_{1} \psi_{N} dV \\
			\int \limits_{D_{e}} \nabla \psi_{2} \psi_{1} dV & \int \limits_{D_{e}} \nabla \psi_{2} \psi_{2} dV & \dots & \int \limits_{D_{e}} \nabla \psi_{2} \psi_{N} dV \\
			\vdots & \vdots & \ddots & \vdots \\
			\int \limits_{D_{e}} \nabla \psi_{N} \psi_{1} dV & \int \limits_{D_{e}} \nabla \psi_{N} \psi_{2} dV & \dots & \int \limits_{D_{e}} \nabla \psi_{N} \psi_{N} dV \\
		\end{bmatrix}
	\end{aligned}
\end{equation}
\begin{equation}
	\label{Eq:29}
	\Theta  = \begin{bmatrix}
		\int \limits_{ \partial D_{e}} \psi_{1} |\mathbf{J}^{s}| d \Gamma  \\
		\int \limits_{\partial D_{e}}  \\psi_{2} |\mathbf{J}^{s}| d \Gamma  \\
		\vdots  \\
		\int \limits_{\partial D_{e}}  \psi_{N} |\mathbf{J}^{s}| d \Gamma \\
	\end{bmatrix}
\end{equation}
\begin{equation}
	\label{Eq:30}
	\Theta^{'}  = \begin{bmatrix}
		\int \limits_{ D_{e}} \psi_{1} |\mathbf{J}^{v}| dV  \\
		\int \limits_{ D_{e}}  \psi_{2} |\mathbf{J}^{v}| dV  \\
		\vdots  \\
		\int \limits_{ D_{e}}  \psi_{N} |\mathbf{J}^{v}| dV \\
	\end{bmatrix}
\end{equation}

In modal DG technique, the preference of orthogonal polynomial functions considerably facilitates the beneficence of the high-order moments of the approximate solutions to the reaction terms associated vector $\Theta^{'}$in Eq. (\ref{Eq:30}). Due to the orthogonal property of the polynomial functions and a simultaneous correlation $\psi_{1}=1$, when the scaled Legendre polynomial functions, $\psi_{N}$, are multiplied by the mapping Jacobian $|\mathbf{J}^{v}|=A_{e}/4$, the integration in the range $[-1,1]$ vanishes for all but the first term.
\begin{equation}
	\label{Eq:31}
	\Theta^{'}  = \begin{bmatrix}
		\int \limits_{ D_{e}} \psi_{1} |\mathbf{J}^{v}| dV  \\
		\int \limits_{ D_{e}}  \psi_{2} |\mathbf{J}^{v}| dV  \\
		\vdots  \\
		\int \limits_{ D_{e}}  \psi_{N} |\mathbf{J}^{v}| dV \\
	\end{bmatrix}
	= \begin{bmatrix}
		4  \\
		0  \\
		\vdots  \\
		0 \\
	\end{bmatrix}
\end{equation}
The reaction term handling is substantially simplified in this unique way, exactly as it is in the first-order ($P^{0}$) case. In short, when the left sides of Eqs. (\ref{Eq:23}) and (\ref{Eq:24}) is derived using the high-order polynomial approximations, the participation of the cell average solutions is dominating the reaction terms in the modal DG construction.

\subsection{Numerical Integration}
\label{Sec:3.4}

The volume and surface integrals appearing in the DG weak formulation (\ref{Eq:23}) are calculated using the Gauss-Legendre quadrature formulas \cite{Signh2018Thesis}. For numerical integration inside the element and over the surface, the quadrature rules need the precise polynomials of degree $2k$ and $2k+1$, respectively. The Gauss-Legendre quadrature of order $N_{q}$ for the reference line and rectangular elements in standard interval $[-1,1]$ is expressed as
\begin{equation}
	\label{Eq:32}
	\begin{aligned}
		\int \limits_{D_{e}} f(x) \, dV & = \mathbf{J}_{e} \cdot \int \limits_{ D_{st}} f(\xi) d \hat{V}  
		=  \mathbf{J}_{e} \cdot \sum_{i=1}^{N_{q}} \omega_{i} f(\xi^{\star}_{i}).\\
		\int \limits_{D_{e}} f(x,y) \, dV & = \mathbf{J}_{e} \cdot \int \limits_{D_{st}} f(\xi,\eta) \, d \hat{V} 
		=  \mathbf{J}_{e} \cdot \sum_{i=1}^{N_{q}} \sum_{j=1}^{N_{q}} \omega_{i} \omega_{j} f(\xi^{\star}_{i},\eta^{\star}_{j}),\\
	\end{aligned}
\end{equation}
In these expressions, $\mathbf{J}_{e}$ denotes the Jacobian of the transformation for the reference element, respectively. 
$\xi^{\star}_{i},\eta^{\star}_{j}$ are the evaluation points on the reference element, and $\omega_{i}, \omega_{j}$ are the weight related with the evaluation points. $N_{q}$ represents the number of evaluation points for the reference cell.

\subsection{Time discretization}
\label{Sec:3.5}

Finally, the spatial discretization of the NCRD systems with the mixed-type DG construction results in a semi-discrete system of ordinary differential equations as described in Eq. (\ref{Eq:26}), which are solved by the following explicit third-order TVD Runge-Kutta scheme \cite{cockburn1998runge}:
\begin{equation}
	\label{Eq:33}
	\begin{aligned}
		\mathbf{U}_{h}^{(1)} & = \mathbf{U}_{h}^{n} + \Delta t \mathbf{L}(U_{h}^{n}),\\
		\mathbf{U}_{h}^{(2)} & = \frac{3}{4} \mathbf{U}_{h}^{n} + \frac{1}{4} \mathbf{U}_{h}^{(1)} + \frac{1}{4} \Delta t  \mathbf{L}(U_{h}^{(1)}),\\
		\mathbf{U}_{h}^{n+1} & = \frac{1}{3} \mathbf{U}_{h}^{n} + \frac{2}{3} \mathbf{U}_{h}^{(2)} + \frac{2}{3} \Delta t  \mathbf{L}(\mathbf{U}_{h}^{(2)}),
	\end{aligned}
\end{equation}
where $\mathbf{L}(U_{h}^{n})$ is the residual of approximate solutions at time $t_n$, and $\Delta t$ is the suitable time-stepping length.

\begin{table}
	\caption{One-dimensional linear RD system: comparison of $L_{\infty}$-error for $u$ with $\mu_{1}=1.0$, $k_{1}=0.1$, $k_{2}=0.01$.}
	\centering
	\begin{tabular}{llllll}
		\hline
		Time & CN-MG \cite{chou2007numerical} & IIF2 \cite{chou2007numerical}  & ECSCM \cite{ersoy2015numerical} & DQM \cite{jiwari2018numerical} &  Present  \\
		\hline
		0.04  & 1.09E-04  & 2.73E-05 & 1.10E-04  & 2.02E-05 &  1.85E-05     \\
		0.02  & 2.67E-05  & 6.40E-06 & 2.84E-05  & 2.07E-05 &  2.76E-06     \\
		0.01  & 6.27E-06  & 1.19E-06 & 7.91E-06  & 2.09E-06 &  2.52E-06     \\
		0.005 & 1.16E-06  & 1.07E-07 & 2.79E-06  & 2.10E-06 &  1.13E-06     \\
		\hline
	\end{tabular}
	\label{Table:1}
\end{table}
\begin{table}
	\caption{One-dimensional linear RD system: error analysis with $\mu_{1}=0.001$, $k_{1}=0.1$, $k_{2}=0.01$ at $t=1$.}
	\centering
	\begin{tabular}{llllll}
		\hline
		$l$ & $N_{x}$ & $L_{2}$-norm & order &  $L_{\infty}$-norm & order     \\
		\hline
		1  & $20$  & $6.751 \times 10^{-2}$ & $-$    & $4.213 \times 10^{-2}$ &  $-$     \\
		& $40$     & $1.804 \times 10^{-2}$ & $1.91$ & $1.105 \times 10^{-2}$ &  $1.85$  \\
		& $80$     & $3.613 \times 10^{-3}$ & $1.87$ & $2.410 \times 10^{-3}$ &  $1.96$  \\
		& $160$    & $2.138 \times 10^{-3}$ & $1.79$ & $1.036 \times 10^{-3}$ &  $2.01$  \\
		2  & $20$  & $1.041 \times 10^{-3}$ & $-$    & $1.248 \times 10^{-3}$ &  $-$     \\
		& $40$     & $1.252 \times 10^{-4}$ & $3.01$ & $1.592 \times 10^{-4}$ &  $2.65$  \\
		& $80$     & $1.732 \times 10^{-5}$ & $3.10$ & $2.007 \times 10^{-5}$ &  $2.75$  \\
		& $160$    & $2.100 \times 10^{-6}$ & $3.02$ & $2.510 \times 10^{-6}$ &  $2.98$  \\
		3  & $20$  & $1.767 \times 10^{-5}$ & $-$    & $2.701 \times 10^{-5}$ &  $-$     \\
		& $40$     & $1.101 \times 10^{-6}$ & $4.02$ & $1.701 \times 10^{-6}$ &  $3.69$  \\
		& $80$     & $7.015 \times 10^{-8}$ & $4.00$ & $3.126 \times 10^{-7}$ &  $3.71$  \\
		& $160$    & $5.252 \times 10^{-9}$ & $4.01$ & $5.125 \times 10^{-8}$ &  $3.92$  \\ 
		\hline
	\end{tabular}
	\label{Table:2}
\end{table}

\section{Results of spatiotemporal pattern formation in NCRD system, and discussion}
\label{Sec:3}

The developed mixed-type DG technique is applied to a variety of the NCRD systems in this section. Subsequently, we elaborate the spatiotemporal pattern formation of several renowned NCRD models, including as linear, Gray-Scott, Brusselator, Isothermal and Schnakenberg, with brief explanation of their natural activities. The spatiotemporal pattern formation of these models have been studied purposely for the same parametric values as used  by previous literature for checking the reliability of the proposed numerical technique. The third-order DG technique was used to accomplish all computations, which were carried out using Visual Studio 2019 and FORTRAN 95 forum. The post-processing work is executed on TECPLOT 360 software.
\begin{figure}
	\centering
	\includegraphics[width=1.0\linewidth]{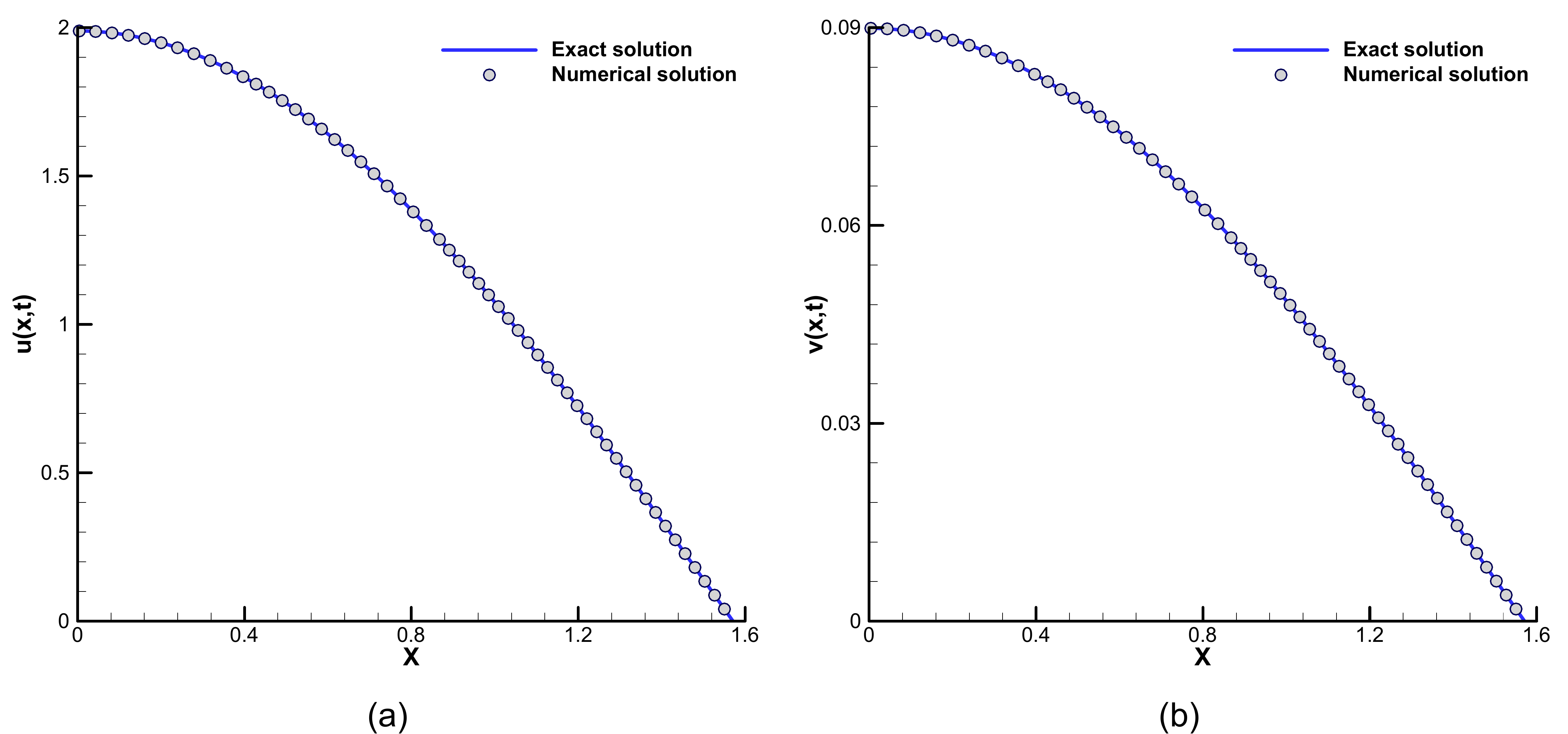}
	\caption{One-dimensional linear RD system: solution patterns for $u$ and $v$ with $\mu_{1}=1.0$, $k_{1}=0.1$, $k_{2}=0.01$.}
	\label{Fig:3}
\end{figure}

\subsection{One-dimensional linear RD system}
\label{Sec.4.1}

As a first problem, we consider one-dimensional linear RD system with domain $\Omega = [0,\pi/2]$ as follows \cite{mittal2016numerical,chou2007numerical}
\begin{equation}
	\label{Eq:34}
	\begin{cases}
		\frac{\partial u}{\partial t} = \mu_{1} \frac{\partial^{2} u}{\partial x^{2}} 
		-k_{1}u +v, & (x,t) \in \Omega \times[0,T],\\
		\frac{\partial v}{\partial t} = \mu_{2} \frac{\partial^{2} v}{\partial x^{2}} -
		k_{2}v, & (x,t) \in \Omega \times[0,T], \\
		u_x(0,t) =u(\pi/2,t) =0,  & t \in \times[0,T], \\
		v_x(0,t) =v(\pi/2,t) =0,  & t \in \times[0,T], 
	\end{cases}
\end{equation}
for some real parameters $k_{1}$ and $k_{2}$. For $\mu_{1}=\mu_{2}$, the exact solution of above system is given by Chou et al. \cite{chou2007numerical} 
\begin{equation}
	\label{Eq:35}
	\begin{cases}
		u(x,t) = \left[e^{-(k_{1} + \mu_{1})t} + e^{-(k_{2} + \mu_{1})t} \right] \cos x, & \\
		v(x,t) =  (k_{1}-k_{2})e^{-(k_{1} + \mu_{1} )t}  \cos x. & \\
	\end{cases}
\end{equation}
The initial conditions is obtained from the true solutions of system (\ref{Eq:35}) by taking $t=0$. We compute the numerical solutions for this RD system with the parameters: $\mu_{1}=1.0$, $k_{1}=0.1$ and $k_{2}=0.01$, which shows a diffusion dominated case. 
$\Delta t =0.001$ is considered for the time discretization, , and simulations are run up to $t=1$. All the computations are done with $N=100$ and the DG scheme with third-order of accuracy. In Table \ref{Table:1}, $L_{\infty}$-norms are estimated for the $u$ variable, and the computed results are compared against the existing numerical results \cite{chou2007numerical,ersoy2015numerical,jiwari2018numerical}. Also, a profile comparison between the true and numerical solutions for $u$ and $v$ at $t=1$ is illustrated in Fig. \ref{Fig:3}. The present numerical results are found quite similar to the true solutions, which demonstrate that the proposed numerical scheme is well efficient to capture the spatiotemporal patterns wave accurately. Furthermore, the order of accuracy of the proposed numerical scheme is also estimated. For this purpose, simulations are run till $t=1.0$, and upto $4^{th}$-order precision by varying the number of cell points $N_{x} (20,40,80,160)$. In Table \ref{Table:2}, the convergence analysis and the order of accuracy has been studied. This investigation shows that the current numerical approach reached the necessary order of accuracy $(l+1)$.

\begin{figure}
	\centering
	\includegraphics[width=1.0\linewidth]{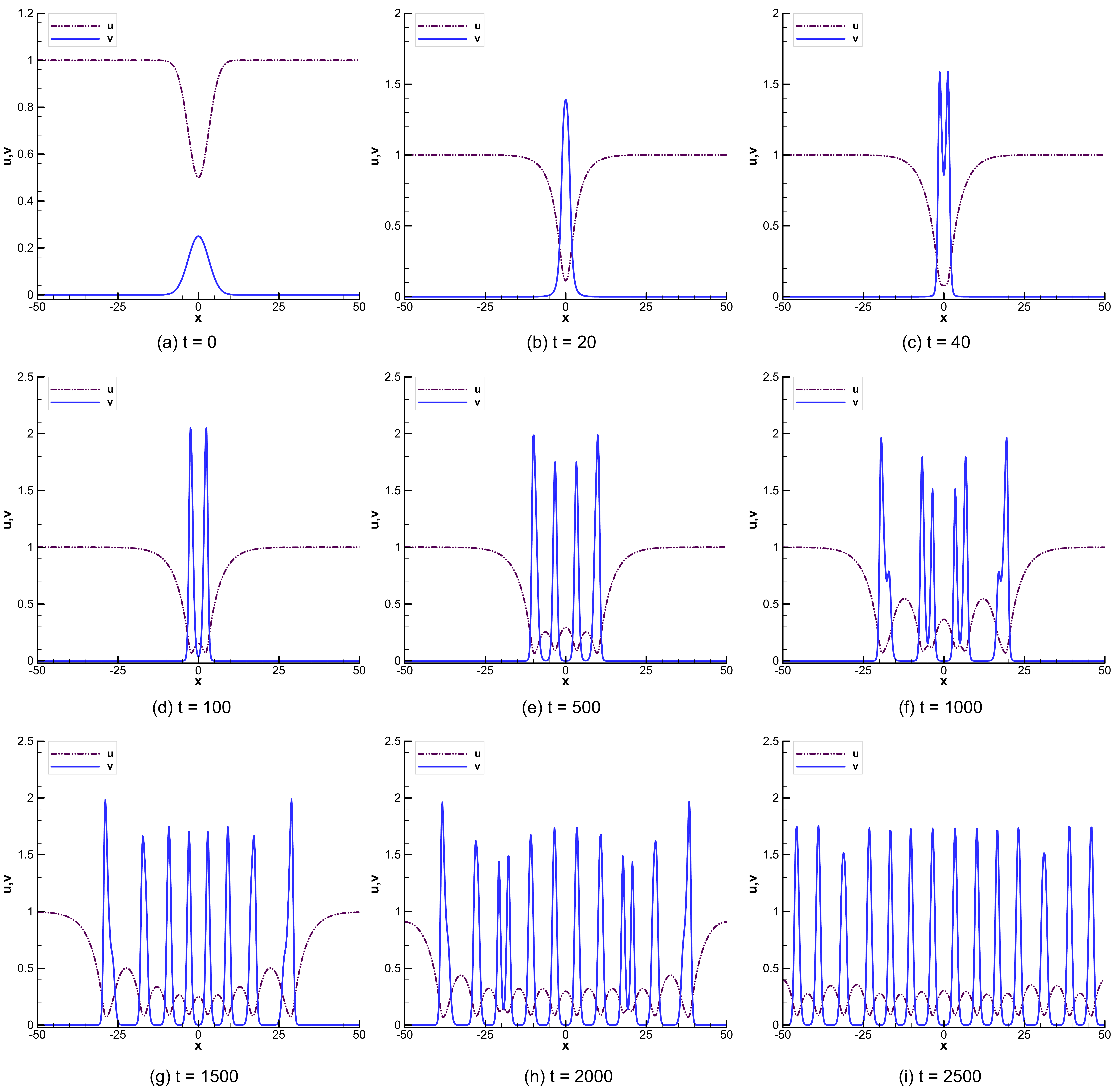}
	\caption{One-dimensional Gray-Scott NCRD system: solution patterns for $u$ and $v$ with $\mu_{1}=1.0$, $\mu_{2}=0.01$, $k_{1}=0.064$ and $k_{2}=0.062$ at different time instants.}
	\label{Fig:4}
\end{figure}
\begin{figure}
	\centering
	\includegraphics[width=1.0\linewidth]{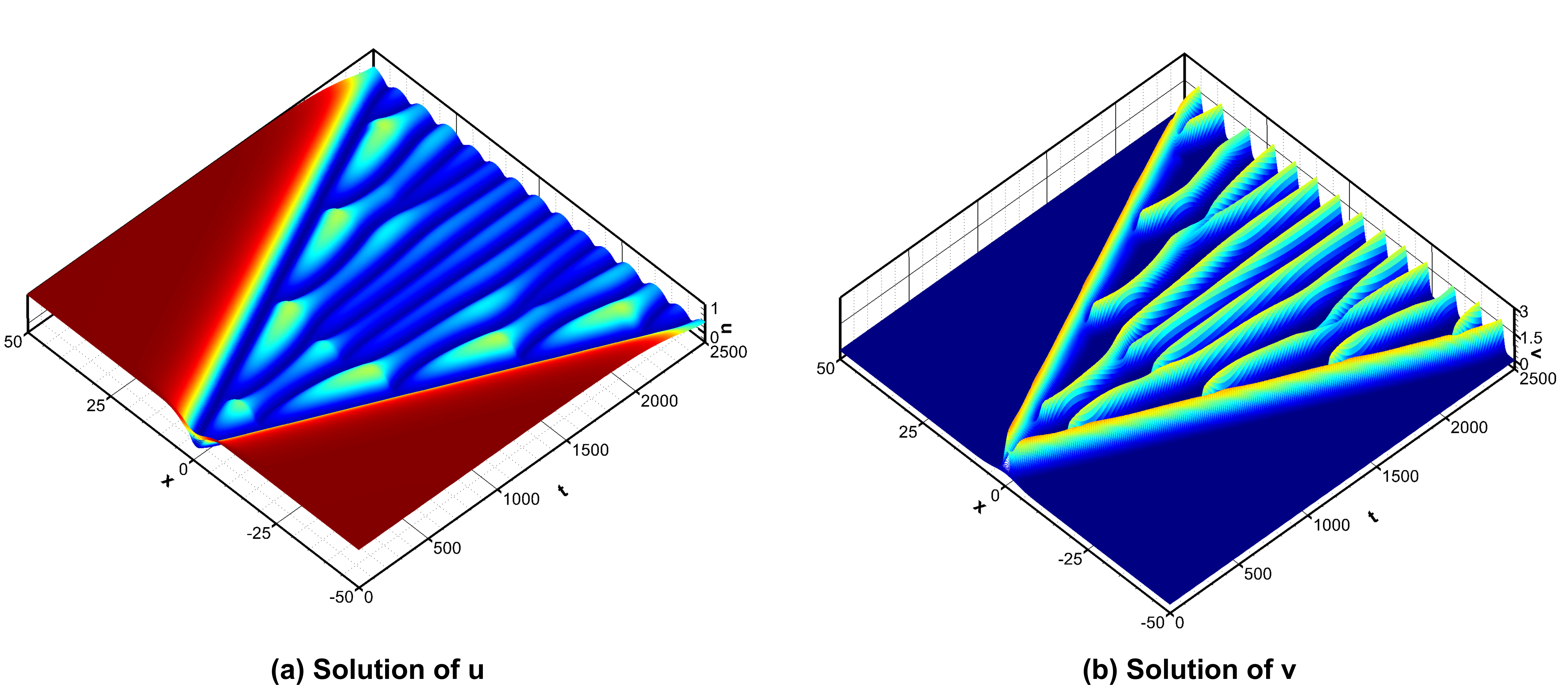}
	\caption{One-dimensional Gray-Scott NCRD system: space-time contours for $u$ and $v$ with $\mu_{1}=1.0$, $\mu_{2}=0.01$, $k_{1}=0.064$ and $k_{2}=0.062$.}
	\label{Fig:5}
\end{figure}

\subsection{One-dimensional Gray-Scott NCRD system}
\label{Sec.4.2}

The Gray-Scott NCRD system, originally proposed by Gray and Scott \cite{gray1984autocatalytic}, describes the various patterns phenomena existing in nature including Turing, butterfly wings, multiple spots and embryo. In this model, the two kinematic-reactions are studied as \cite{pearson1993complex} 
\[\begin{aligned}
	U + 2V \rightarrow 3V, \\
	V \rightarrow P,
\end{aligned} \]
where $U$ and $V$ are the chemical species, and $P$ denotes an inert outcome of these species. Both kinematic-reactions are irreversible, therefore, the densities of $U$ and $V$ remain consistent. Both chemical species are being removed in the process and the resulting dimensionless Gray-Scott RD system with domain $\Omega = [-50,50]$ becomes as
\begin{equation}
	\label{Eq:36}
	\begin{cases}
		\frac{\partial u}{\partial t} = \mu_{1} \frac{\partial^{2} u}{\partial x^{2}} 
		-uv^{2} +k_{1}(1-u), & (x,t) \in \Omega \times[0,T],\\
		\frac{\partial v}{\partial t} = \mu_{2} \frac{\partial^{2} v}{\partial x^{2}} + uv^{2} - (k_{1}+k_{2})v, & (x,t) \in \Omega \times[0,T], \\
		u(-50,t) = u(50,t) = 1,  & t \in \times[0,T], \\
		v(-50,t) = v(50,t) = 0,  & t \in \times[0,T]. 
	\end{cases}
\end{equation}
For this system, we consider the following initial conditions
\begin{equation}
	\label{Eq:37}
	\begin{aligned}
		u(x,0) & = 1-\frac{1}{2}sin^{100} \left(\pi \frac{(x-50)}{100}\right), \\
		v(x,0) & = \frac{1}{4}sin^{100} \left(\pi \frac{(x-50)}{100}\right).
	\end{aligned}
\end{equation}
In aforementioned expression, $u$ and $v$ denote the the concentrations of chemical species $U$ and $V$, respectively.  $\mu_{1}$, $\mu_{2}$ denote the diffusion rates; $k_{1}$ represents the feed rate of $u$ into the system and $k_{2}$ represents the reaction rate of second reaction. 
For this system, we evaluate the solutions with the following selected parameters: $\mu_{1}=1.0$, $\mu_{2}=0.01$, $k_{1}=0.064$ and $k_{2}=0.062$. All the numerical experiments are carried out with $400$ elements on the spatial domain $\Omega=[-50,50]$ with $\Delta t=0.01 $ and the simulations are run untill $t=2500$ so that the  stationary wave patterns are obtained. The growth, separation and replication of the solution patterns are observed in Figs. \ref{Fig:4}-\ref{Fig:5}. The wave patterns are continuously growing, splitting and replicating themselves until the entire computational domain is covered by these waves. During the pulse-splitting process, some new pulses are developed in the rear part of the prevailing pulses as demonstrated in Fig. \ref{Fig:5}. The obtained spatiotemporal patterns are very compatible with the existing literature \cite{mittal2016numerical,jiwari2017numerical,tok2019wave}.

\begin{figure}
	\centering
	\includegraphics[width=1.0\linewidth]{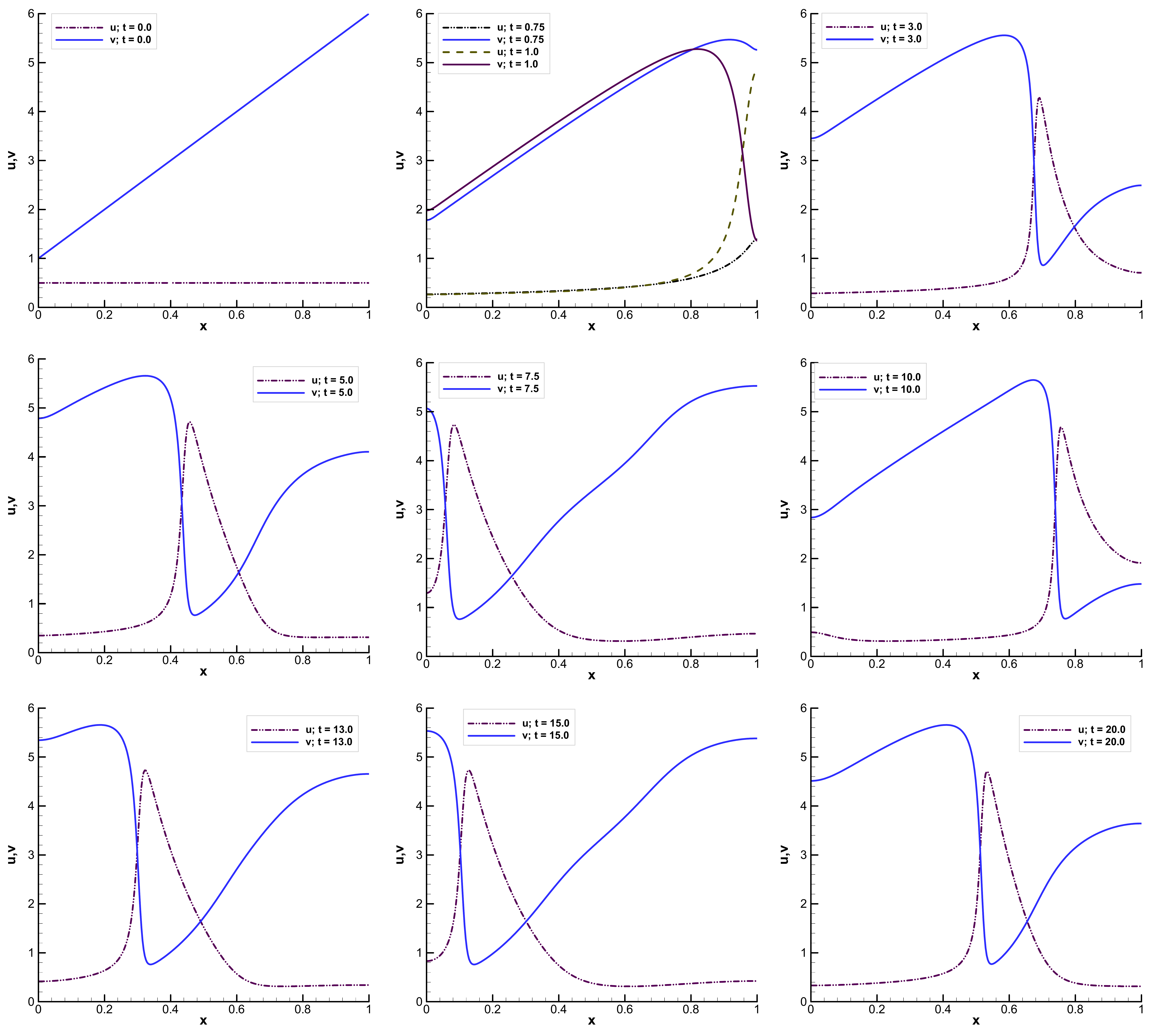}
	\caption{One-dimensional Brusselator NCRD system: spatiotemporal profiles for  $u$ and $v$ with $\mu_{1}=\mu_{2}=10^{-4}, k_{1}=1.0, k_{2}=3.4$ at different time instants.}
	\label{Fig:6}
\end{figure}
\begin{figure}
	\centering
	\includegraphics[width=1.0\linewidth]{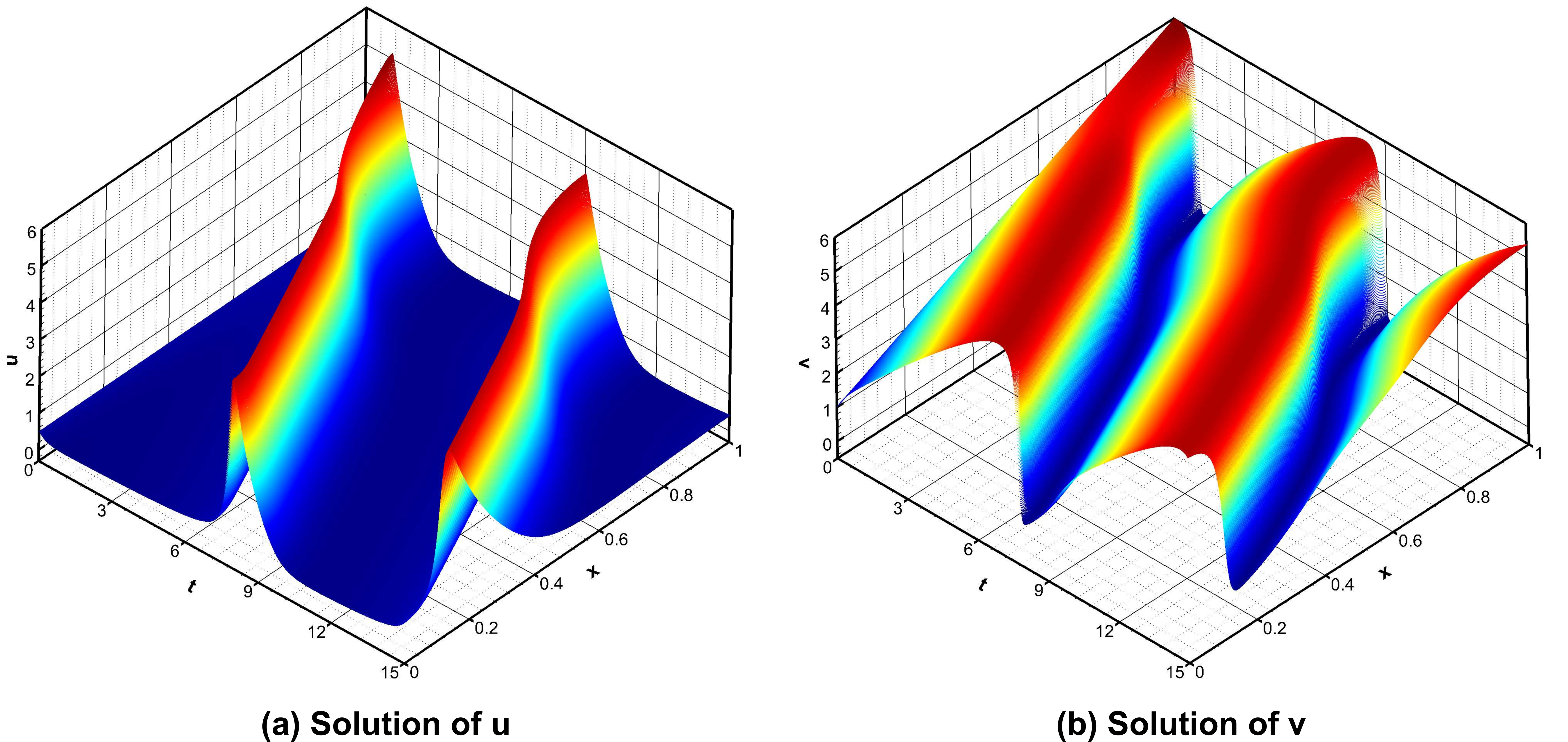}
	\caption{One-dimensional Brusselator NCRD system: spatiotemporal contours for $u$ and $v$ with $\mu_{1}=\mu_{2}=10^{-4}, k_{1}=1.0, k_{2}=3.4$.}
	\label{Fig:7}
\end{figure}

\subsection{One-dimensional Brusselator NCRD system}
\label{Sec.4.3}

The Brusselator NCRD system is crucial in real-world physical issues such as turing pattern creation on animal skin, ozone synthesis from atomic oxygen, enzyme reactions, laser and plasma physics due to multiple mode coupling. This system was originated by Prigogine and R. Lefever in 1968 \cite{lefever1968dissipative,prigogine1968symmetry}. It uses highly nonlinear oscillations to illustrates the chemical RD process by the following tri-molecular chemical system
\cite{herschkowitz1972localized,lefever1971chemical,Nicolis1977self,twizell1999second}:
\[\begin{aligned}
	k_{1} & \rightarrow U, \\
	k_{2} + U &\rightarrow V + D, \\
	2U + V &\rightarrow 3U, \\
	U &\rightarrow E, \\
\end{aligned} \]
where $k_{1}$ and $k_{2}$ are the input chemicals, $D$ and $E$ are  the output chemicals, and $U$ and $V$ represent the intermediates. The second reaction represents the bi-molecular reaction, while, the third one denotes the autocatalytic tri-molecular reaction. The resulting dimensionless Brusselator reaction-diffusion system with domain $\Omega=[0,1]$ is defined as \cite{zegeling2004adaptive}
\begin{equation}
	\label{Eq:38}
	\begin{cases}
		\frac{\partial u}{\partial t} = \mu_{1} \frac{\partial^{2} u}{\partial x^{2}} 
		+ u^{2}v - (1+k_{2})u + k_{1}, & (x,t) \in \Omega \times[0,T],\\
		\frac{\partial v}{\partial t} = \mu_{2} \frac{\partial^{2} v}{\partial x^{2}} - u^{2}v + k_{2} u, & (x,t) \in \Omega \times[0,T], \\
		u_{x}(0,t) = u_{x}(1,t) =0,  & t \in \times[0,T], \\
		v_{x}(0,t) = v_{x}(1,t) =0,  & t \in \times[0,T], 
	\end{cases}
\end{equation}
with the following initial conditions
\begin{equation}
	\label{Eq:39}
	\begin{cases}
		u(x,0) = 0.5, \\
		v(x,0) = 1 + 5x.
	\end{cases}
\end{equation}
For the numerical experiments, we choose the following parameters: $\mu_{1}=\mu_{2}=10^{-4}$, $k_{1}=1.0$, $k_{2}=3.4$, based on previous studies \cite{zegeling2004adaptive}. The computational work has been done with $200$ elements on the spatial domain $\Omega=[0,1]$ and the used time-step is $\Delta t=0.01$ and simulations are run until $t=20$. Figures \ref{Fig:6}-\ref{Fig:7} show the spatiotemporal pattern formation of the Brusselator NCRD system. These patterns of the Brusselator system are found very similar to previous studies \cite{zegeling2004adaptive,mittal2016numerical,jiwari2017numerical}. In Fig. \ref{Fig:6}, the physical behaviour of $u$ and $v$ solution profiles during the molecular interactions are observed at different time instants. At $t=0$, these profiles are found high on the right side of the domain. As time increases, these spatiotemporal profiles move towards left side of the  domain.

\begin{figure}
	\centering
	\includegraphics[width=1.0\linewidth]{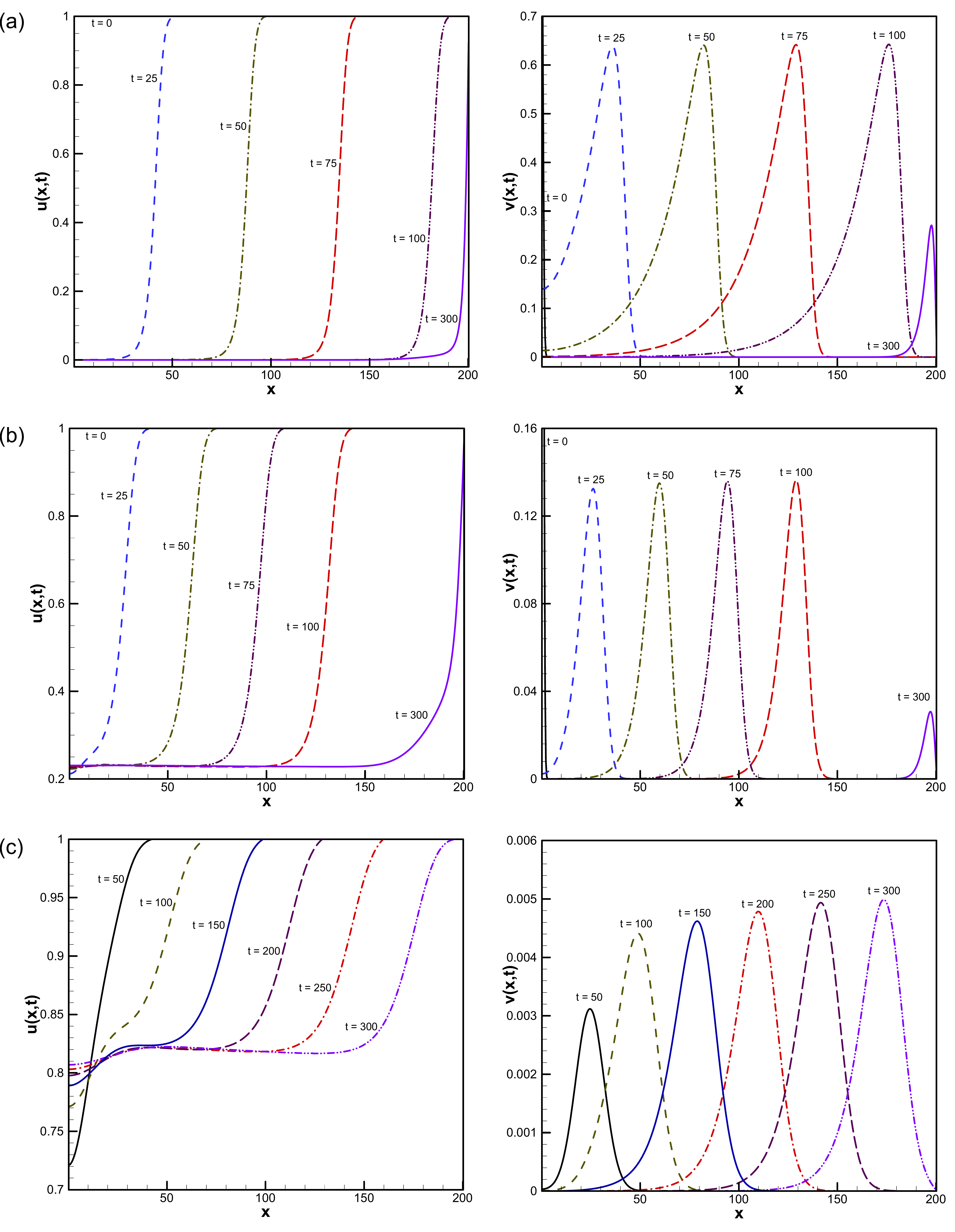}
	\caption{One-dimensional isothermal chemical NCRD system: spatiotemporal profiles for $u$ and $v$ with $\mu_{1}=\mu_{2}=1$, and (a) $k_{1}=0.1$, (b) $k_{1}=0.5$, (c) $k_{1}=0.9$ at different time instants.}
	\label{Fig:8}
\end{figure}
\begin{figure}
	\centering
	\includegraphics[width=1.0\linewidth]{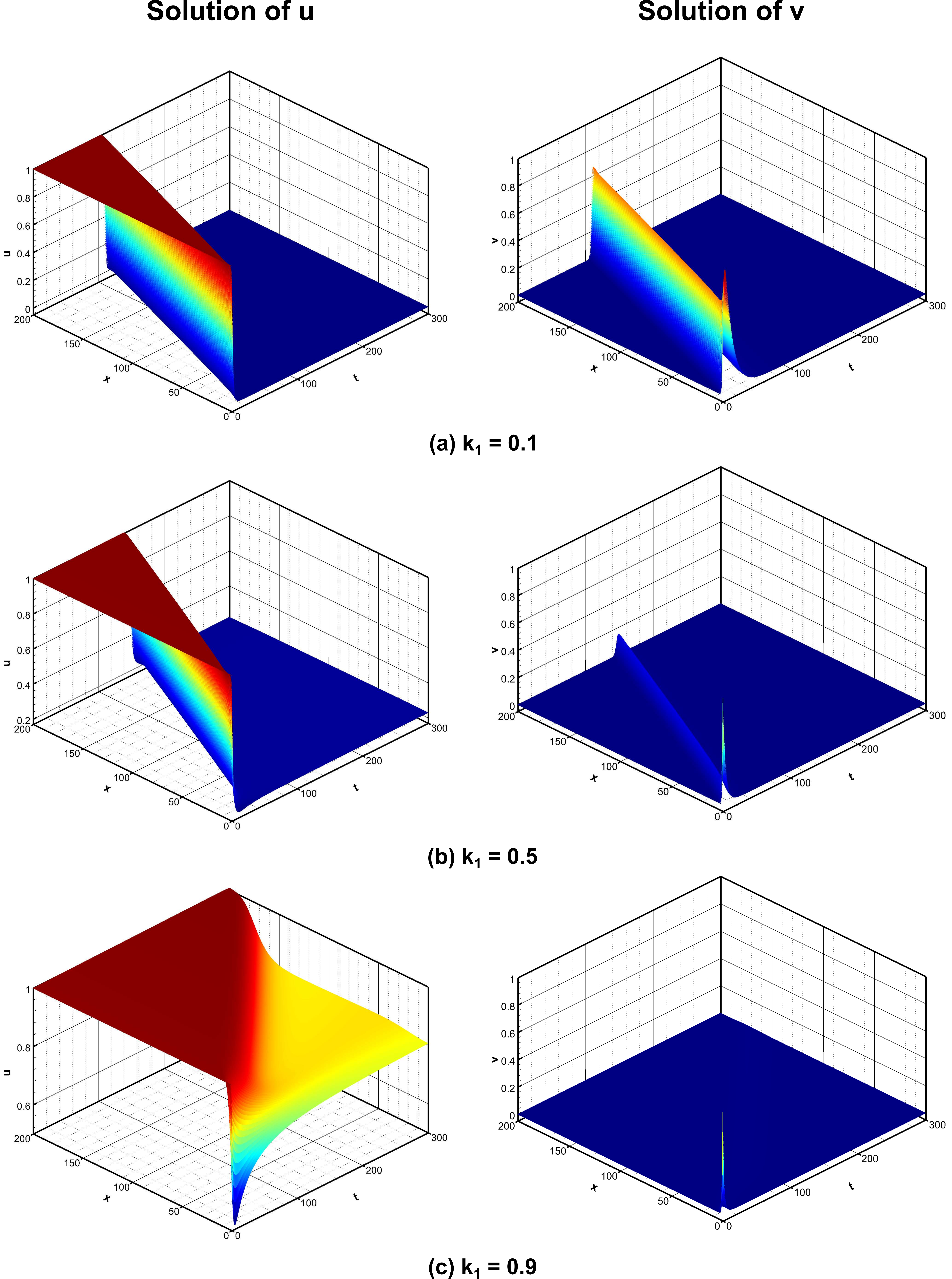}
	\caption{One-dimensional isothermal chemical NCRD system: spatiotemporal contours for $u$ and $v$ with $\mu_{1}=\mu_{2}=1$, and (a) $k_{1}=0.1$, (b) $k_{1}=0.5$, (c) $k_{1}=0.9$.}
	\label{Fig:9}
\end{figure}

\subsection{One-dimensional isothermal chemical NCRD system}
\label{Sec.4.4}

The isothermal chemical NCRD system was developed with auto-catalytic reactions \cite{merkin1990development,garcia1996linearized} 
\[\begin{aligned}
	U + 2V & \rightarrow 3V, \\
	V + V &\rightarrow P,  
\end{aligned} \]
where $U$ and $V$ illustrates the chemical reactants, and $P$ denotes the output reactant. 
The resulting dimensionless isothermal chemical NCRD system with $\Omega=[0,200]$ is defined as \cite{garcia1996linearized} 
\begin{equation}
	\label{Eq:40}
	\begin{cases}
		\frac{\partial u}{\partial t} = \mu_{1} \frac{\partial^{2} u}{\partial x^{2}} 
		-uv, & (x,t) \in \Omega \times[0,T],\\
		\frac{\partial v}{\partial t} = \mu_{2} \frac{\partial^{2} v}{\partial x^{2}} + uv  -k_{1} v, & (x,t) \in \Omega \times[0,T], \\
		u_{x}(0,t) =u_{x}(200,t) =0,  & t \in \times[0,T], \\
		v_{x}(0,t) =v_{x}(200,t) =0,  & t \in \times[0,T].
	\end{cases}
\end{equation}
For this system, we adopt the following initial conditions
\begin{equation}
	\label{Eq:41}
	\begin{cases}
		u(x,0) = 1, \\
		v(x,0) = e^{-x^2}.
	\end{cases}
\end{equation}
In an isothermal chemical NCRD system, $k_{1}>0$ represents a dimensionless variable that expresses the intensity of the crumble step to the autocatalytic fashioning phase. It was demonstrated using this system that almost-isothermal flames emerging from quadratic branching in the carbon sulphide oxygen reaction could be characterized using basic cubic autocatalysis \cite{merkin1990development}. Based on the previous studies \cite{garcia1996linearized,mittal2016numerical,jiwari2017numerical,ersoy2015numerical}, the numerical experiments are carried out with the parameters: $\mu_{1}=\mu_{2}=1$, and $k_{1}=0.1$, $0.5$, $0.9$. For this NCRD system, we performed the numerical simulations with $300$ elements over the  region $\Omega=[0,200]$ with $\Delta t=0.01$, and the simulations are run till time $t=300$. The impact of different $k_{1}$ values on the spatiotemporal pattern formations of $u$ and $v$ can be illustrated in Figs. \ref{Fig:8}-\ref{Fig:9}. For all $f$ values, the spatiotemporal profiles for $u$ and $v$ solutions transport in the direction of right side of the domain with time increasing as shown in Fig. \ref{Fig:7}, and the concentration of $u$ remains constant towards the boundary, whereas $v$ decreases to zero.
Further, the physical behaviour of the spatiotemporal profiles can be also revealed by the space-time plots as shown in Fig. \ref{Fig:9}. The spatiotemporal patterns obtained by the present DG scheme are quite similar to the previous studies \cite{garcia1996linearized,mittal2016numerical,jiwari2017numerical,ersoy2015numerical}.

\begin{figure}
	\centering
	\includegraphics[width=1.0\linewidth]{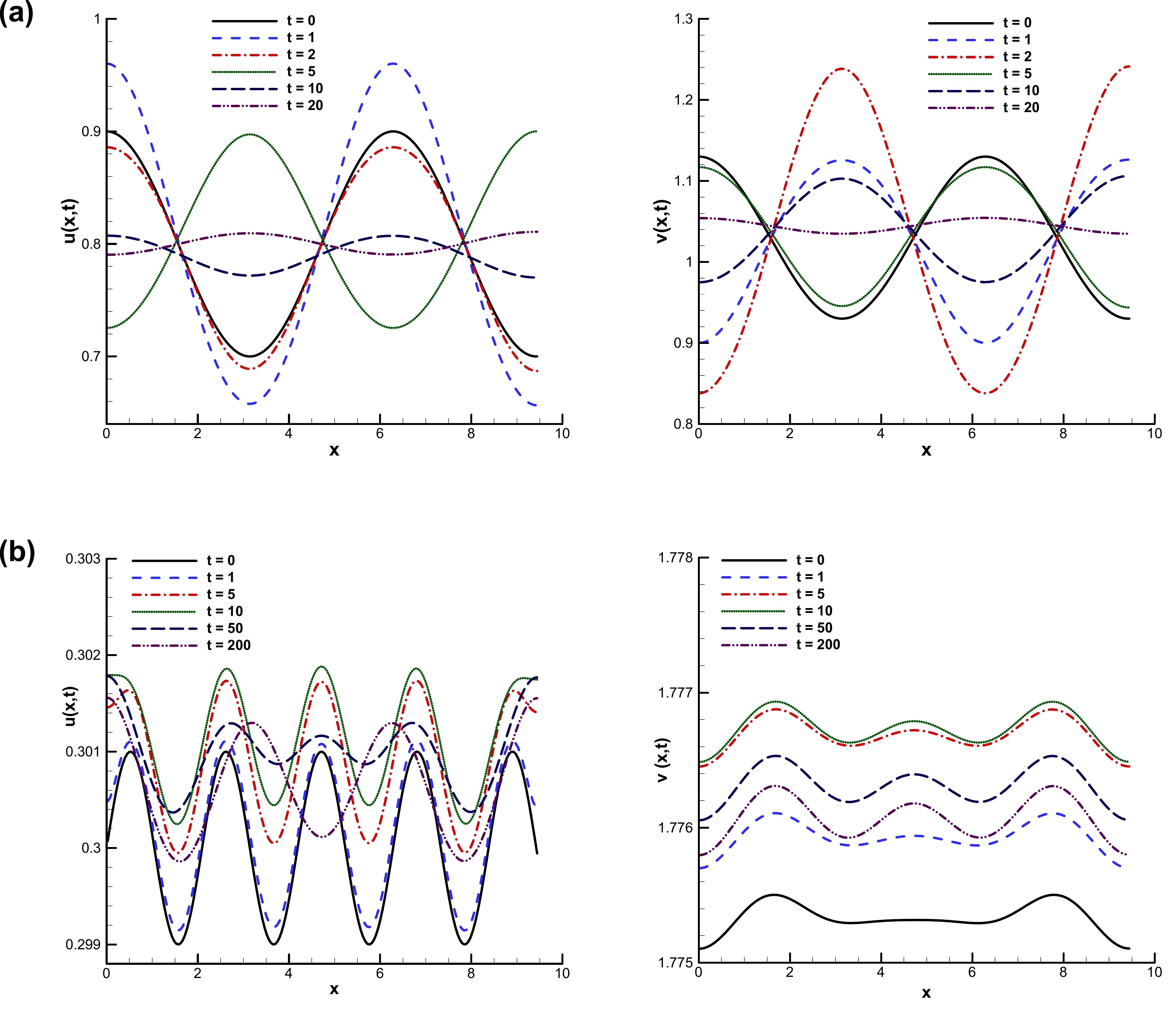}
	\caption{One-dimensional Schnakenberg NCRD system: spatiotemporal profiles for  $u$ and $v$ with (a) $\mu_{1}=0.2, \mu_{2}=0.1$, $k_{1}=0.14$, $k_{2}=0.66$ with $\text{I.Cs. (i)}$, and (b) $\mu_{1}=0.01, \mu_{2}=1.0$, $k_{1}=0.14$, $k_{2}=0.16$ with $\text{I.Cs. (ii)}$ at different time instants.}
	\label{Fig:10}
\end{figure}
\begin{figure}
	\centering
	\includegraphics[width=1.0\linewidth]{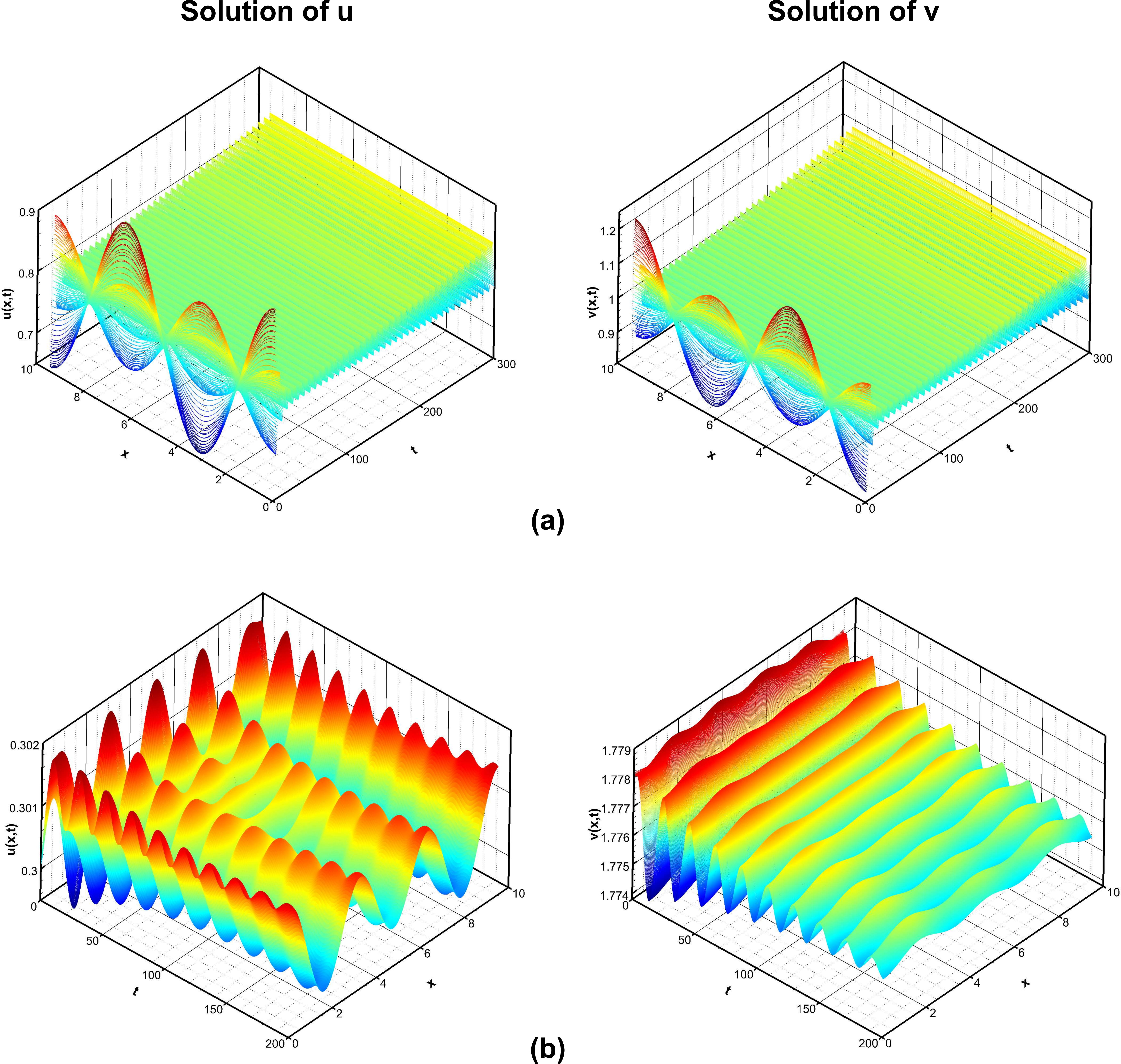}
	\caption{One-dimensional Schnakenberg NCRD system: spatiotemporal contours for $u$ and $v$ with (a) $\mu_{1}=0.2, \mu_{2}=0.1$, $k_{1}=0.14$, $k_{2}=0.66$ with $\text{I.Cs. (i)}$, and (b) $\mu_{1}=0.01, \mu_{2}=1.0$, $k_{1}=0.14$, $k_{2}=0.16$ with $\text{I.Cs. (ii)}$.}
	\label{Fig:11}
\end{figure}

\subsection{One-dimensional Schnakenberg NCRD system}
\label{Sec.4.5}

The Schnakenberg NCRD system \cite{schnakenberg1979simple} is a fundamental model that describes an autocatalytic chemical reaction with possible oscillatory behaviour. The tri-molecular reactions between two chemical products $U,V$ and two chemical source $A,B$ are formulated as:  
\[\begin{aligned}
	A & \rightleftharpoons U, \\
	B &\rightarrow V, \\ 
	2U + V &\rightarrow 3U. 
\end{aligned} \]
A system of two nonlinear RD equations for the concentrations $u$ and $v$ of the chemical products $U$ and $V$ can be obtained through the law of mass action.
A Schnakenberg NCRD system in dimensionless form with domain $\Omega=[0,3 \pi]$ can defined as \cite{Murray1989Biology,ghergu2011nonlinear,jiwari2017numerical}: 
\begin{equation}
	\label{Eq:42}
	\begin{cases}
		\frac{\partial u}{\partial t} = \mu_{1} \frac{\partial^{2} u}{\partial x^{2}} 
		+u^{2}v - u + k_{1}, & (x,t) \in \Omega \times[0,T],\\
		\frac{\partial v}{\partial t} = \mu_{2} \frac{\partial^{2} v}{\partial x^{2}} -u^{2}v  + k_{2}, & (x,t) \in \Omega \times[0,T], \\
		u_{x}(0,t) =u_{x}(3 \pi,t) =0,  & t \in \times[0,T], \\
		v_{x}(0,t) =v_{x}(3 \pi,t) =0,  & t \in \times[0,T], 
	\end{cases}
\end{equation}
with the following two initial conditions:
\begin{equation}
	\label{Eq:43}
	\text{I.Cs. (i)} =
	\begin{cases}
		u(x,0) = 0.8 + 0.1 \cos x, \\
		v(x,0) = 1.03 + 0.1 \cos x. \\
	\end{cases}
	\text{I.Cs. (ii)} =
	\begin{cases}
		u(x,0) = 0.3 + 0.001 \sin 3x, \\
		v(x,0) = 1.778 + 0.001 \cos 2x. \\
	\end{cases}
\end{equation}
Based on the study of Jiwari et al. \cite{jiwari2017numerical}, we have selected the following parameters for computational simulations: (a) $\mu_{1}=0.2, \mu_{2}=0.1$,  $k_{1}=0.14$, $k_{2}=0.66$ with initial conditions $(\text{I.Cs. (i)})$; and (b) $\mu_{1}=0.01, \mu_{2}=1.0$,  $k_{1}=0.14$, $k_{2}=0.16$ with initial conditions $(\text{I.Cs. (ii)})$. 
For this system, we performed the numerical simulations with $300$ elements over the  region $\Omega=[0,3 \pi]$ with a time-step $\Delta t=0.001$ and DG solver is run to time $t=300$. The spatiotemporal patterns with two different initial conditions and different parameters are illustrated in Figs. \ref{Fig:9}-\ref{Fig:10}. The numerical results shows that solution converges to a spatially homogeneous periodic orbit with the first initial condition (\ref{Eq:43}),  whereas with the second initial condition, the solutions display a spatially periodic spatiotemporal pattern with wave length $\pi$. Further, the physical behaviour of the spatiotemporal patterns can be also revealed by the space-time plots as shown in Fig. \ref{Fig:11}. The spatiotemporal patterns obtained by the present DG scheme are quite similar to the results of Jiwari et al. \cite{jiwari2017numerical}. It can also be observed that the spatiotemporal patterns of the system strictly depend upon the initial conditions and parameters, as shown in Figs. \ref{Fig:10}-\ref{Fig:11}, which are drastically different.

\begin{figure} [hbt!]
	\centering
	\includegraphics[width=1.0\linewidth]{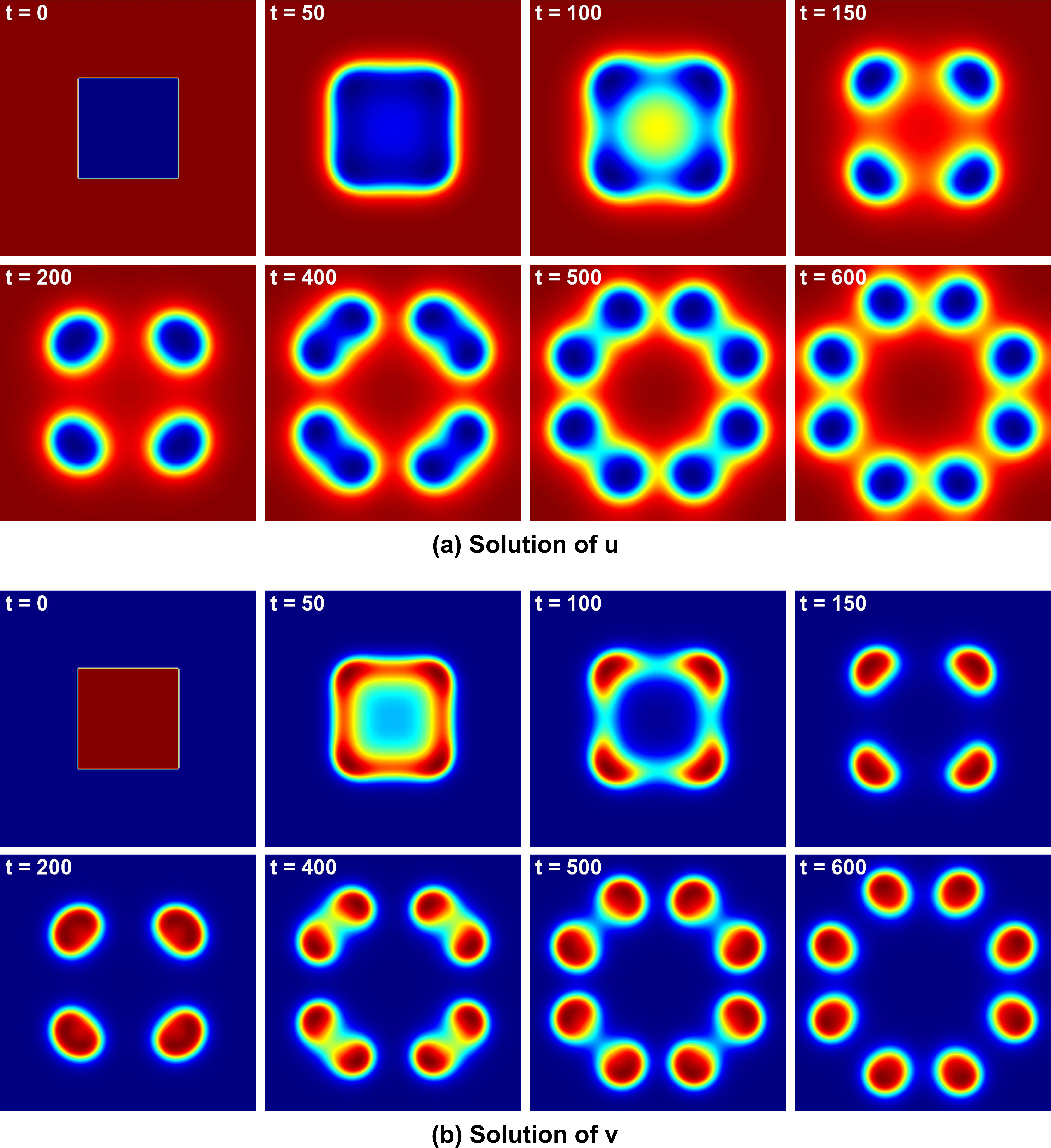}
	\caption{Two-dimensional Gray-Scott NCRD system: spatiotemporal contours for (a) solution of $u$, and (b) solution of $v$ with the parameters $\mu_{1}=8 \times 10^{-5}$, $\mu_{2}=4 \times 10^{-5}$, $k_{1}=0.024$, $k_{2}=0.06$ at different time instants.}
	\label{Fig:12} 
\end{figure}

\subsection{Two-dimensional Gray-Scott NCRD system}
\label{Sec.4.6}

For capturing the spatiotemporal pattern formation, the two-dimensional NCRD system is more obvious than one-dimensional system. Therefore, we consider a two-dimensional Gray-Scott NCRD system in this test case as follows \cite{zegeling2004adaptive}:
\begin{equation}
	\label{Eq:44}
	\begin{cases}
		\frac{\partial u}{\partial t} = \mu_{1} \left[\frac{\partial^{2} u}{\partial x^{2}} + \frac{\partial^{2} u}{\partial y^{2}}\right] -uv^{2} + k_{1}(1-u), & (x,y,t) \in \Omega \times[0,T],\\
		\frac{\partial v}{\partial t} = \mu_{2} \left[\frac{\partial^{2} v}{\partial x^{2}} + \frac{\partial^{2} v}{\partial y^{2}}\right]+ uv^{2} - (k_{1} + k_{2})v, & (x,y,t) \in \Omega \times[0,T]. \\
	\end{cases}
\end{equation}
For this system, we consider two block functions as the initial condition defined in the domain $\Omega =[0,1] \times [0,1]$ as
\begin{equation}
	\label{Eq:45}
	\begin{aligned}
		u(x,y,0) & = \begin{cases} 0.5 \;\;  & \text{if } 0.3 \leq x \leq 0.7 \;\; \text{and} \;\;  0.3 \leq y \leq 0.7, \\
			1              & \text{otherwise},
		\end{cases}\\
		v(x,y,0) & = \begin{cases} 0.25 \;\;  & \text{if } 0.3 \leq x \leq 0.7 \;\; \text{and} \;\;  0.3 \leq y \leq 0.7, \\
			0              & \text{otherwise}.
		\end{cases}\\       	
	\end{aligned}
\end{equation}
For this NCRD system, the two-dimensional domain is partitioned with the grid points $151 \times 151$, and the Dirichlet conditions are used for the boundary implementation. For simulations, the time step of $\Delta t = 10^{-3}$ is used. The following parameters are considered for computations: $\mu_{1}=8 \times 10^{-5}$, $\mu_{2}=4 \times 10^{-5}$, $k_{1}=0.024$, $k_{2}=0.06$. Figure \ref{Fig:12} illustrates the spatiotemporal contours for $u$ and $v$ at distinct time intervals $t=0,50,100,150,300,400,500,600,800,$ and 1000. The two-dimensional split proceeding of the original block functions into four spots and then eight sports in terms of the first and second concentration components. From the results, it is found that the spatiotemporal patterns obtained by the present DG scheme are quite similar to the results of Zegeling and Kok \cite{zegeling2004adaptive}.

\begin{figure} [hbt!]
	\centering
	\includegraphics[width=1.0\linewidth]{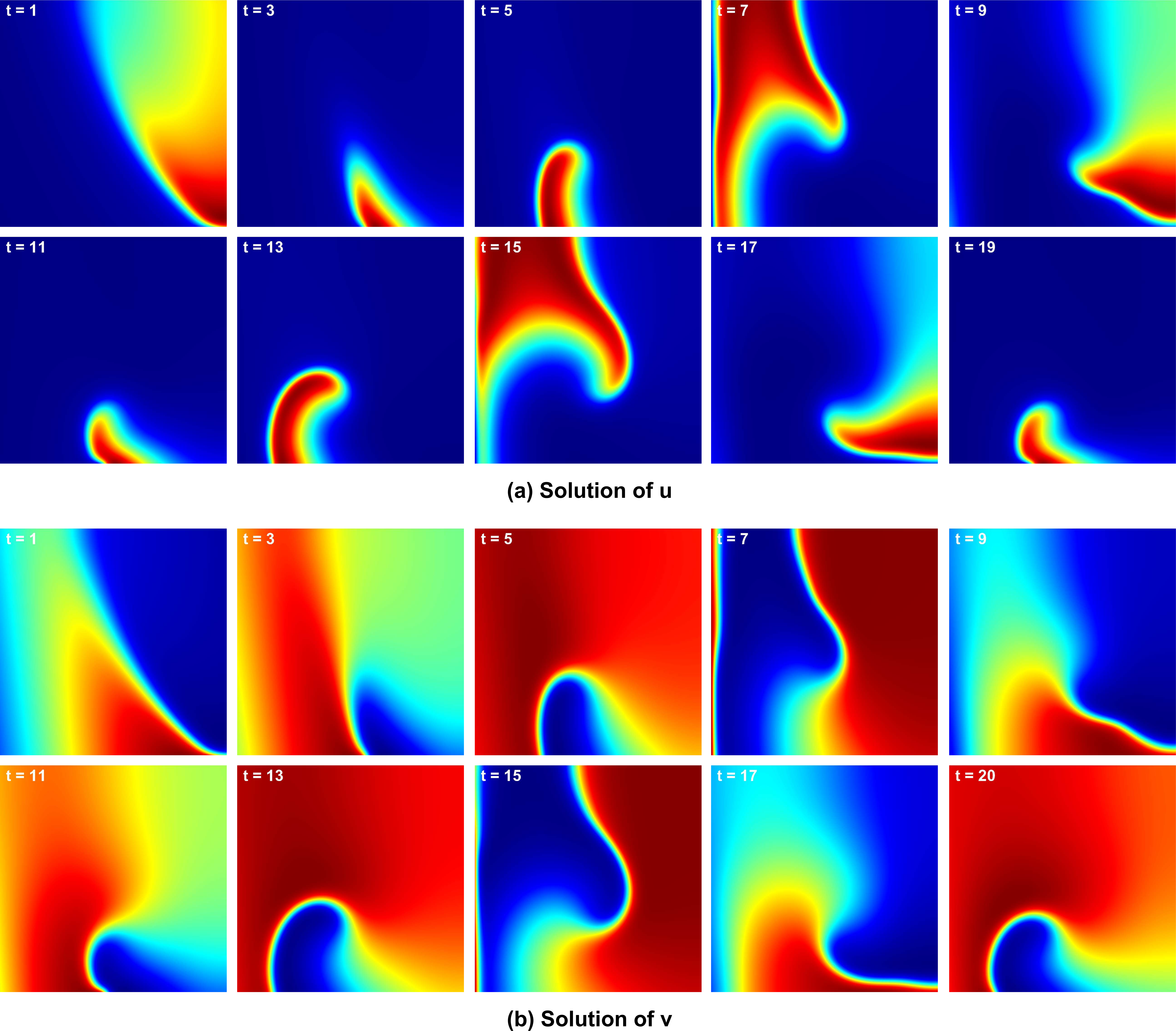}
	\caption{Two-dimensional Brusselator NCRD system: spatiotemporal contours for (a) solution of $u$, and (b) solution of $v$ with the parameters $\mu_{1}=\mu_{2}=2.0\times 10^{-3}$, $k_{1}=1$, $k_{2}=3.4$ at different time instants.}
	\label{Fig:13} 
\end{figure}

\subsection{Two-dimensional Brusselator NCRD system}
\label{Sec.4.7}

Finally, a two-dimensional Brusselator NCRD system is examined as follows
\cite{zegeling2004adaptive,ADOMIAN19951}:
\begin{equation}
	\label{Eq:46}
	\begin{cases}
		\frac{\partial u}{\partial t} = \mu_{1} \left[\frac{\partial^{2} u}{\partial x^{2}} + \frac{\partial^{2} u}{\partial y^{2}}\right] + uv^{2} - (1+k_{1}) u + k_{2}, & (x,y,t) \in \Omega \times[0,T],\\
		\frac{\partial v}{\partial t} = \mu_{2} \left[\frac{\partial^{2} v}{\partial x^{2}} + \frac{\partial^{2} v}{\partial y^{2}}\right] - uv^{2} + k_{1} u, & (x,y,t) \in \Omega \times[0,T]. \\
	\end{cases}
\end{equation}
For this NCRD system, the following initial conditions are considered along with Neumann boundary conditions in the computational $\Omega =[0,1] \times [0,1]$:
\begin{equation}
	\label{Eq:47}
	\begin{aligned}
		u(x,y,0) & = 0.5 + y, \\
		v(x,y,0) & = 1 + 5 x. \\
	\end{aligned}
\end{equation}
Numerical simulations are performed with the following parameters: $\mu_{1}=\mu_{2}=2.0\times 10^{-3}$, $k_{1}=1$, and $k_{2}=3.4$.  Figure  \ref{Fig:13} illustrates the spatiotemporal contours for the concentrations $u$ and $v$ at distinct time intervals. From the results, it is obvious that due to $k_{2}>1 + k_{1}$,  the solutions do not converge, and they oscillate on a regular basis. The spatiotemporal patterns travel from $y-$ axis to $x-$ axis, as shown in the contour plots. These plots like those in Zegeling and Kok \cite{zegeling2004adaptive} provide the similar behaviors as the 2D Brusselator NCRD system.

\section{Concluding remarks}
\label{Sec:4}

The nonlinear coupled reaction-diffusion (NCRD) systems are important in the formation of spatiotemporal patterns in many scientific and engineering fields. In this work, we apply a mixed-type modal discontinuous Galerkin (DG) approach to one- and two- dimensional NCRD systems such as linear, Gray-Scott, Brusselator, isothermal, and Schnakenberg models to yield the spatiotemporal patterns. These models essentially represent a variety of complicated natural spatiotemporal patterns such as spots, spot replication, stripes, hexagons, and so on. In this approach, a mixed-type formulation is presented to address the second-order derivatives emerging in the diffusion terms. For spatial discretization, hierarchical modal basis functions premised on the orthogonal scaled Legendre polynomials are used.  Moreover, a novel reaction term treatment is proposed for the NCRD system, demonstrating an intrinsic feature of the new DG scheme and preventing erroneous solutions due to extremely nonlinear reaction terms. The proposed approach reduced the NCRD system into a set of ordinary differential equations in time, which are solved by an explicit third-order TVD Runge-Kutta scheme. We considered seven different test cases from the literature and gave simulation results for their spatiotemporal patterns in order to validate the numerical DG method. The spatiotemporal patterns generated with the present approach are very comparable to those found in the literature. The major goal of this study is to look into the formation of spatiotemporal patterns in one-dimensional and two-dimensional NCRD systems. Additional forces, like as electric and magnetic fields, play a prominent role in the creation of spatiotemporal patterns in multi-dimensional NCRD systems. The cross-diffusion rate is also expected to have a substantial impact on the development of the spatiotemporal pattern in NCRD systems. In this regard, future work will examine the effects of electromagnetic fields, particularly the cross-diffusion rate, on spatiotemporal pattern forms in multi-dimensional varied NCRD systems.

\section*{Acknowledgements}

S.S. and M.B. would like to thank Nanyang Technological University Singapore for their financial support under the NAP-SUG award.




\end{document}